\documentclass[reprint, twocolumn, amsfonts, amsmath, amssymb, superscriptaddress, prl]{revtex4-2}

\usepackage{graphicx}
\usepackage{dcolumn}
\usepackage{bm}
\usepackage[normalem]{ulem}
\usepackage{mathtools}
\usepackage[dvipsnames]{xcolor}
\usepackage[linkcolor=Blue,citecolor=Blue,urlcolor=Blue,colorlinks=true]{hyperref}

\begin{document}

\title{Enzyme-enriched condensates show self-propulsion, positioning, and coexistence}
\author{Leonardo Demarchi}
\thanks{These authors contributed equally to this work.}
\altaffiliation[Present address: ]{Sorbonne Université, CNRS, Institut de Biologie Paris-Seine (IBPS), Laboratoire Jean Perrin (LJP), F-75005 Paris, France}
\affiliation{Arnold Sommerfeld Center for Theoretical Physics and Center for NanoScience, Department of Physics, Ludwig-Maximilians-Universit\"at M\"unchen, Theresienstra\ss e 37, D-80333 M\"unchen, Germany}
\author{Andriy Goychuk}
\thanks{These authors contributed equally to this work.}
\altaffiliation[Present address: ]{Institute for Medical Engineering and Science, Massachusetts Institute of Technology, Cambridge, MA 02139, United States}
\email{andriy.goychuk@gmail.com}
\affiliation{Arnold Sommerfeld Center for Theoretical Physics and Center for NanoScience, Department of Physics, Ludwig-Maximilians-Universit\"at M\"unchen, Theresienstra\ss e 37, D-80333 M\"unchen, Germany}
\author{Ivan Maryshev}
\affiliation{Arnold Sommerfeld Center for Theoretical Physics and Center for NanoScience, Department of Physics, Ludwig-Maximilians-Universit\"at M\"unchen, Theresienstra\ss e 37, D-80333 M\"unchen, Germany}
\author{Erwin Frey}
\email{frey@lmu.de}
\affiliation{Arnold Sommerfeld Center for Theoretical Physics and Center for NanoScience, Department of Physics, Ludwig-Maximilians-Universit\"at M\"unchen, Theresienstra\ss e 37, D-80333 M\"unchen, Germany}
\affiliation{Max Planck School Matter to Life, Hofgartenstraße 8, D-80539 M\"unchen, Germany}


\begin{abstract}
Enzyme-enriched condensates can organize the spatial distribution of their substrates by catalyzing non-equilibrium reactions.
Conversely, an inhomogeneous substrate distribution induces enzyme fluxes through substrate-enzyme interactions.
We find that condensates move towards the center of a confining domain when this feedback is weak. 
Above a feedback threshold, they exhibit self-propulsion, leading to oscillatory dynamics.
Moreover, catalysis-driven enzyme fluxes can lead to interrupted coarsening, resulting in equidistant condensate positioning, and to condensate division.
\end{abstract}

\maketitle

Liquid-liquid phase separation in living cells can lead to the formation of biomolecular condensates that aid intracellular organization~\cite{Brangwynne2009, Hyman2014, Banani2017, Lyon2020, Alberti2019, Shin2017, Choi2020}.
These condensates have different functions such as compartmentalization of reactions~\cite{Lyon2020}, buffering of molecules~\cite{Klosin2020}, and midcell localization during cell division~\cite{Schumacher2017}. 
However, in a thermal equilibrium system, the liquids will completely segregate through a coarsening process (Ostwald ripening)~\cite{Ostwald1900, LifshitzSlyozov1961, Wagner1961, Bray1994}.
To arrest this process, the system must be brought out of equilibrium by supplying energy, e.g., via fuel-driven chemical reactions.
This has been shown to lead to `active droplet' systems that exhibit a wealth of novel phenomena not encountered in thermal equilibrium~\cite{Shin2017, Banani2017, Weber2019}.

Previous studies have considered systems with a continuous turnover of condensate (droplet) material by chemical reactions~\cite{Weber2019, Glotzer1994, Glotzer1995, Christensen1996, Carati1997, Zwicker2015, Lamorgese2016, Zwicker2016, Wurtz2018, Li2020, Kirschbaum2021}.
The resulting material fluxes lead to multi-droplet coexistence~\cite{Carati1997, Zwicker2015, Wurtz2018, Li2020, Kirschbaum2021} and droplet division~\cite{Zwicker2016}.
Here, we study a different class of systems where conserved enzymes spontaneously phase separate, or localize to an existing condensate~\cite{supplement}.
These enzymes then regulate reactions among other molecules, by transiently binding substrate and catalyzing its conversion into product via a lower activation barrier.
For example, in the bacterium \emph{Myxococcus xanthus}, a PomXY cluster (moving on the nucleoid) regulates the cycling of PomZ between two conformations~\cite{Schumacher2017, Bergeler2018, Kober2019, Hanauer2021}.
We show that such substrate turnover and the resulting enzyme fluxes lead to condensate self-propulsion, positioning, interrupted coarsening, and condensate division.
Interestingly, previous studies have shown that liquid droplets can self-propel on a surface through active stresses~\cite{Joanny2012, Khoromskaia2015, Whitfield2016, Kree2017, Yoshinaga2019, Trinschek2020}, altering their wetting properties~\cite{Thiele2004, John2005}, or in viscous fluids through Marangoni flows~\cite{MichelinReview2022}.
In contrast, in our case, condensate motion is driven by the bulk interactions between the various chemical species and does not require surfaces or hydrodynamic coupling.

While condensates might consist of several components, here we focus on the enzyme concentration $c(\mathbf{x},t)$.
To describe the dynamics of liquid-liquid phase separation, we take the Cahn-Hilliard (CH) equation as a starting point with the following chemical potential~\cite{Cahn1961}: ${\mu_0 (c) = - r (c-\tilde c) + u (c-\tilde c)^3 - \kappa \boldsymbol\nabla^2 c}$.
This chemical potential ${\mu_0 (c) = \delta {\cal F}[c] / \delta{c}}$ corresponds to the Ginzburg-Landau free energy functional ${{\cal F}[c]}$ for a symmetric binary mixture with the critical density $\tilde c$ and phenomenological parameters $r$, $u$, and $\kappa$; in particular, the control parameter $r$ measures the distance from the critical point~\cite{Bray1994}. 
The enzymes interact with substrates and products, which are present at concentrations $s(\mathbf{x},t)$ and $p(\mathbf{x},t)$. 
These couplings, quantified by the Flory-Huggins (FH) parameters $\chi_s$ and $\chi_p$, modify the local chemical potential of enzymes.
Assuming that the particle currents are proportional to gradients in the chemical potential~\cite{DeGroot2013, Balian2006}, the enzyme dynamics is given by 
\begin{equation}
\label{eq:enzymes}
    \partial_t c ({\bf x},t) 
    = 
    \boldsymbol\nabla \cdot 
    \big[ 
    M \, c \, \boldsymbol\nabla 
    \big( 
    \mu_0 (c) + \chi_s s + \chi_p p 
    \big) 
    \big]
    \, ,
\end{equation}
where $M$ denotes a mobility and the term in the square brackets is the enzyme flux $\mathbf{j}(\mathbf{x},t)$.
This \textit{gradient dynamics} leads to a  gradual minimization of the free energy functional from which it is derived~\cite{supplement}, a hallmark of systems close to thermal equilibrium.
Analogously, the thermodynamic fluxes of substrates and products can be derived from the same free energy functional~\cite{supplement}.

Active systems, however, exhibit processes that break detailed balance in protein reaction networks~\cite{ReviewHalatek2018, ReviewBurkart2022}.
For example, in the conversion of nucleoside triphosphatases (NTPases) between an NDP-bound (``product'') and a less stable NTP-bound (``substrate'') state, abundance of NTP (``fuel'') in solution may shift the equilibrium toward the latter~\cite{ZwickerReview2022} and replenish substrate with the net rate $k_2 \, p$.
Transient binding of an enzyme (NTPase-activating protein, NAP) to substrate can, by lowering the activation barrier of hydrolysis, kinetically select the fuel-independent reaction pathway and replenish product with the net rate $k_1  c \, s$, which follows from the law of mass action.
We assume that these separate reaction pathways are far from their respective equilibria, so that we can disregard thermodynamic constraints~\cite{DeGroot2013, Cates2019} and treat the rate constants $k_{1,2}$ as independent parameters.
Hence, we write for the dynamics of the substrate and the product:
\begin{subequations}
\label{eq:system}
\begin{align}
\label{eq:substrates}
    \partial_t s 
    &= 
    \boldsymbol\nabla \cdot 
    \left( D \, \boldsymbol\nabla s + \Lambda \, s \, \chi_s \boldsymbol\nabla c  
    \right) - k_1 \, c \, s + k_2 \, p \,, \\
    \label{eq:products}
    \partial_t p 
    &= 
    \boldsymbol\nabla \cdot 
    \left( D \, \boldsymbol\nabla p + \Lambda \, p\, \chi_p \boldsymbol \nabla c  
    \right) + k_1 \, c \, s - k_2 \, p \,.
\end{align}
\end{subequations}
Catalytic substrate turnover is preceded by enzyme binding, suggesting effective pairwise attraction, ${\chi_s < 0}$.
Converting substrate into product reduces its affinity for the enzymes, ${\chi_s < \chi_p}$, leading to unbinding.
Note that in Eq.~\eqref{eq:system} we have taken the liberty of formally decoupling the diffusion coefficient $D$ from the mobility $\Lambda$, thus introducing a further source of far from equilibrium dynamics by breaking the fluctuation-dissipation relation valid for thermal equilibrium systems.  
In the present context, this is the Einstein-Smoluchowsky relation ${D=\Lambda \, k_B T}$~\cite{Frey2005}. 
With the aim of simplifying the analysis, in the present work we consider ${\Lambda=0}$, an approximation that is valid for weak FH parameters $\chi_{s,p}$ and will be addressed elsewhere~\cite{DropletsLongPaper}.

\begin{figure}[!b]
\includegraphics{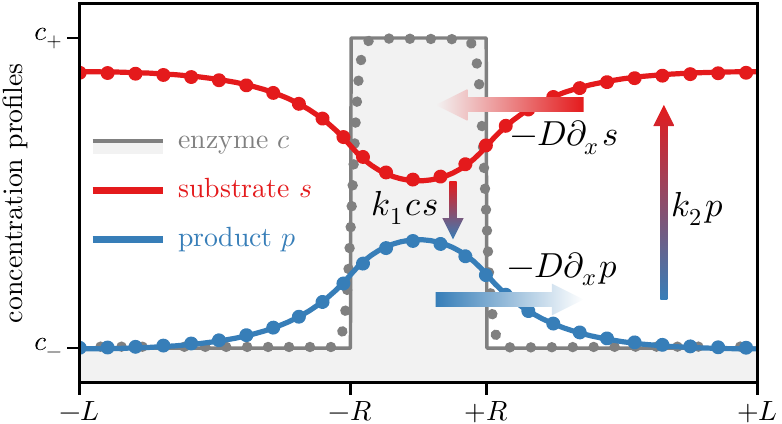}
\caption{
\label{fig:stationary_droplet} 
Steady state concentration profiles for a 1d droplet with no-flux boundary conditions at ${x=\pm L}$.
Arrows indicate reactive (vertical) and diffusive (horizontal) fluxes.
The analytical solutions in the sharp interface approximation (lines) match our simulations (dots).
We use $c_+$ as reference concentration and define the characteristic time ${\tau_0 \coloneqq k_2^{-1}}$, diffusion length in the absence of enzymes ${l_0 \coloneqq \sqrt{D/k_2}}$, and reference energy ${\epsilon_0 \coloneqq r c_+}$.
The remaining parameters, ${c_- = 0.1 c_+}$, ${w = 0.1 l_0}$, ${R = l_0}$, ${L = 5 l_0}$, ${M = 100 D/\epsilon_0}$, ${k_1 = k_2/c_+}$, ${\chi_s = -0.05 r}$, ${\chi_p = -0.01 r}$, and ${s+p = c_+}$, are fixed for all figures unless stated otherwise.
}
\end{figure}

For our initial exploration of the dynamics, we consider a finite-sized domain $[-L,L]$ in a one-dimensional (1d) geometry with no-flux boundary conditions at ${x=\pm L}$.
A droplet then corresponds to a plateau with a high enzyme concentration, surrounded by an enzyme-poor phase [Fig.~\ref{fig:stationary_droplet}].
If the width of the interface between these phases, ${w=\sqrt{2\kappa/r}}$, is much smaller than all other length scales, one can use a sharp interface approximation with piecewise constant concentration $c$.
For our analysis, we consider weak interactions, ${|\chi_{s} s| + |\chi_{p} p| \ll r \, (c_+ - c_-)}$, and therefore approximate the enzyme concentrations in the two phases by their equilibrium values, ${c_\pm = \tilde c \pm \sqrt{r/u}}$.

First, consider a \textit{stationary} droplet, where a closed analytic solution of Eqs.~\eqref{eq:enzymes}~and~\eqref{eq:system} can be obtained~\cite{supplement}.
The stationary state is maintained by a balance of reactive and diffusive fluxes [Fig.~\ref{fig:stationary_droplet}]: In the droplet, the enzymes catalyze the conversion of substrate to product, consuming the former and accumulating the latter.
Diffusive fluxes in turn replenish substrate in the droplet while expelling product, resulting in concentration gradients over the  characteristic diffusion lengths ${l_\pm = \sqrt{D/(k_1 c_\pm + k_2)}}$ inside and outside the droplet, respectively.
This leads to cyclic diffusive and reactive fluxes such that time-reversal symmetry is broken and one has a reaction-driven non-equilibrium steady state.

\begin{figure}[bt]
\includegraphics{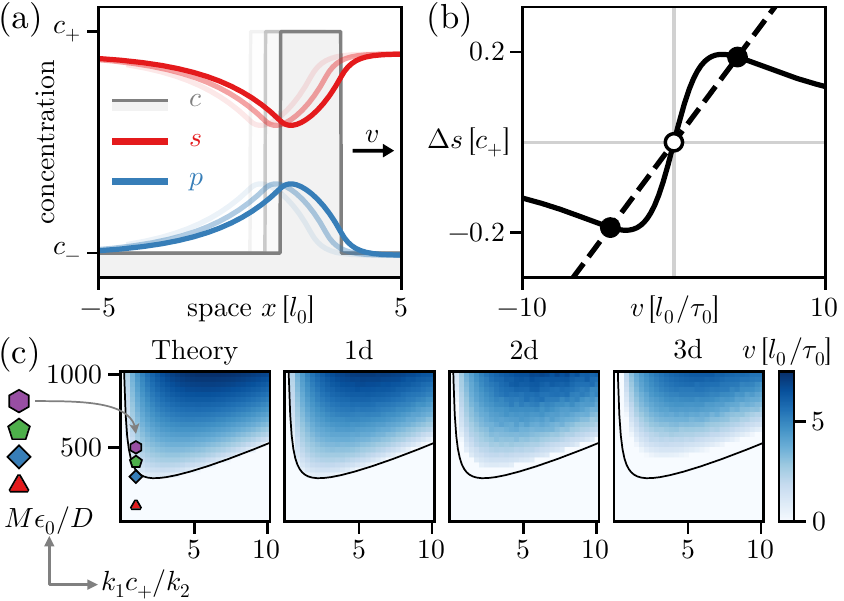}
\caption{\label{fig:instability} 
Self-propulsion instability.  
(a) Analytical profiles for a droplet moving with velocity ${v = 2 \, l_0 / \tau_0}$.
Lighter colors indicate earlier times. 
(b) Graphical analysis of the self-consistency relation~\eqref{eq:self_consistency} for ${M \epsilon_0 / D = 1000}$.
The solid curve indicates the substrate imbalance $\Delta s(v)$, while the slope of the dashed line corresponds to the right-hand side of inequality~\eqref{eq:selfpropulsion_condition}.
Stationary droplets correspond to unstable solutions (empty circle), while self-propelling droplets are stable (filled circles).
(c) Theoretical prediction and simulation results for the self-propulsion velocity (color scale) with $M$ and $k_1$ as free parameters. 
The solid black lines indicate the critical mobility $M^*$ (for the explicit closed expression see~\cite{supplement}). 
}
\end{figure}

If there is an appreciable difference in substrate or product concentration between the two droplet interfaces (henceforth referred to as ``imbalance''), this generally results in a chemical potential gradient that can drive droplet motion through a net flux of enzymes [Eq.~\eqref{eq:enzymes}].
Indeed, using finite element (FEM) simulations of Eqs.~\eqref{eq:enzymes}~and~\eqref{eq:system}, we find a broad parameter regime with ballistic droplet motion [Fig.~\ref{fig:instability}c, Video~1].
To analytically determine the conditions for the onset of this \emph{self-propulsion instability}, we next study the sharp-interface limit of a single 1d droplet in an infinite domain.

Similar to the analysis of Fisher waves~\cite{FISHER1937}, we solve Eq.~\eqref{eq:system} in the reference frame of a moving droplet to obtain the concentration profiles of substrates and products, and then use Eq.~\eqref{eq:enzymes} to derive a self-consistency relation for the droplet velocity $v$.
Specifically, the continuity equation~\eqref{eq:enzymes} implies that the moving steady-state enzyme profile, ${c(z)}$ with ${z \coloneqq x - v t}$ and constant velocity $v$, can only be maintained if ${\partial_z j(z) = v \, \partial_z c(z)}$ holds at all times. 
The local flux of enzymes is therefore given by ${j(z) = v \, [c(z) - c_-]}$, and vanishes in the far field where all concentrations become homogeneous~\cite{supplement}.
The concentration of enzymes in the droplet is enriched by ${\Delta c = c_+ \,{-}\, c_-}$ with respect to the far-field value ${c(\pm\infty) = c_-}$.
While the enzyme flux is driven by the \emph{local} chemical potential and concentration gradients [Eq.~\eqref{eq:enzymes}], integrating the flux over the droplet domain $[-R,R]$ yields an expression that depends only on the values at the droplet boundaries.
In the sharp-interface limit, the $\mu_0$-term becomes mirror-symmetric relative to the droplet center and hence doesn't contribute to the self-consistency relation for the droplet velocity:
\begin{equation}
    2 R \, \Delta c \, v 
    = 
    - M c_+ 
    \big[
    \chi_s \Delta s(v) + \chi_p \Delta p(v)
    \big] 
    \, .
    \label{eq:self_consistency}
\end{equation}
Here, ${\Delta s(v) = s(R) - s(-R)}$ is the substrate concentration imbalance between the two opposite sides of the droplet, with an analogous expression $\Delta p(v)$ for the product. 
This result quantifies how asymmetric substrate and product concentration profiles drive droplet motion. 

Using the closed analytic expressions for the substrate concentration profiles~\cite{supplement}, shown in Fig.~\ref{fig:instability}a,  one can graphically solve the self-consistency relation~\eqref{eq:self_consistency} for the droplet velocity $v$ [Fig.~\ref{fig:instability}b].
Specifically, for ${\Lambda = 0}$, where the total concentration of substrates and products is constant [Eqs.~\eqref{eq:system}], the right-hand side of Eq.~\eqref{eq:self_consistency} simplifies to ${M c_+ (\chi_p - \chi_s) \Delta s(v)}$.
Based on the graphical form of $\Delta s (v)$, a non-vanishing solution to the self-consistency relation for the velocity $v$ exists if [Fig.~\ref{fig:instability}b]: 
\begin{equation}
    \frac{\partial}{\partial v}\Delta s (v)\Big|_{v=0} 
    > 
    \frac{2 R \Delta c}{M c_+ (\chi_p - \chi_s)} 
    \, .
    \label{eq:selfpropulsion_condition}
\end{equation}
Thus, traveling wave solutions emerge if, for example, the mobility $M$ or the difference between the FH parameters ${\chi_p - \chi_s}$ are sufficiently large. 
Then, enzymes are pulled more towards substrates than products, so that enzymatic substrate depletion can induce a chemophoretic effect.
This theoretical analysis quantitatively explains the onset of the self-propulsion instability that we observed in our simulations; see Fig.~\ref{fig:instability}c for a comparison.

\begin{figure}[t]
\includegraphics{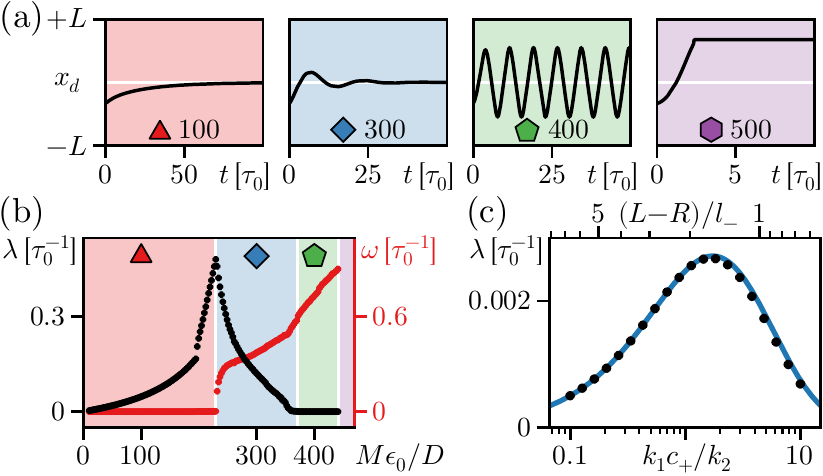}
\caption{\label{fig:positioning} 
Self-centering and oscillations.
(a) Droplet center trajectories in simulations for different values of $M$; symbols indicate the position in the diagram shown in Fig.~\ref{fig:instability}c.
The droplet center is initially at ${x_d(0) = -l_0}$ in a domain of size $L = 3 \, l_0$.
(b) Decay rate (black dots) and frequency (red dots) as functions of $M$, obtained by fitting the respective droplet trajectories.
Red triangle indicates overdamped regime.
(c) Relaxation rate $\lambda$ as a function of $k_1$ (bottom axis), for ${M \epsilon_0 / D = 10}$ and ${x_d(0) = -0.3 l_0}$.
Top axis relates the domain size to the length scale of the concentration profiles.
The analytical predictions (blue line) in the quasi-steady-state approximation~\cite{supplement} match our simulations (black dots).
}
\end{figure}

Droplet movement is driven by asymmetries in the concentration profiles of substrates and products. 
Droplets induce such asymmetries autonomously during self-propulsion, but also near impermeable domain boundaries. 
Figure~\ref{fig:positioning}a and Video~2 show the results of FEM simulations in a closed domain for four characteristic values of the mobility $M$. 
Below the self-propulsion threshold $M^*$, the droplet exhibits an overdamped relaxation toward the domain center (red), where the concentration profiles become symmetric. 
Thus, the impermeable domain boundaries effectively repel the droplet, due to substrate depletion and product enrichment within a range $l_-$. 
When increasing $M^*$, there is a transition from overdamped to underdamped oscillatory relaxation (blue), where the relaxation rate $\lambda$ has a maximum at critical damping [Fig.~\ref{fig:positioning}b], similar to a damped harmonic oscillator. 
Above the self-propulsion threshold $M^*$, droplets autonomously accelerate to a terminal velocity $v$ [Fig.~\ref{fig:instability}c]. 
Instead of droplet self-centering, one then observes oscillations (green) with frequency ${\omega \approx v / (L-l_-)}$, where the domain boundaries cause droplets to slow down and reverse. 
Droplets with strong self-propulsion (purple) can overcome this repulsion and attach to the boundary.

To elucidate how droplet self-centering depends on the reaction rates, we analyzed the overdamped regime, in which droplet motion is much slower than the relaxation of the substrate and product concentration profiles.
This time scale separation allows to solve Eqs.~\eqref{eq:system} analytically using a quasi-stationary approximation, where the substrate and product concentration profiles are in steady state with the droplet center $x_d(t)$ considered as slowly varying.
The obtained steady-state profiles are asymmetric when the droplet is not centered in the domain, if and only if the characteristic length $l_+$ is neither much larger nor much smaller than the droplet size $R$ (lest both droplet interfaces have equal concentrations).
These asymmetric concentration profiles induce droplet motion towards the domain center, see Eq.~\eqref{eq:self_consistency}, with a velocity $v(x_d)$ that we linearize as a function of the distance to the domain center~\cite{supplement}, ${|x_d| \ll l_-}$.
The resulting approximation for the relaxation rate $\lambda$ agrees well with our simulations [Fig.~\ref{fig:positioning}c], and demonstrates that droplet self-centering is fastest for a finite value of $k_1 c_+/k_2$.
Moreover, our analysis and simulations show that droplet self-centering proceeds fastest when the distance between the droplet interface and the domain boundary is comparable to the range of repulsion, ${L-R \sim l_-}$ [Fig.~\ref{fig:positioning}c].

%
A state with multiple droplets cannot be stable in a thermodynamic system.
Instead, a coarsening process driven by interfacial energy minimization takes place, causing smaller droplets to shrink and larger droplets to grow until there is complete phase separation~\cite{LifshitzSlyozov1961, Wagner1961, Bray1994}.
As our system is out of equilibrium, it can result in a stable coexistence of multiple droplets.
Specifically for the system we are considering, larger droplets have a larger enzymatic activity and hence consume more substrate, leading to a reduced substrate concentration at their interfaces.
This results in a gradient of substrate (and product) in the low-concentration phase between droplets of different sizes, thereby transporting enzymes from the larger to the smaller droplets and thus counteracting the coarsening process [Eq.~\eqref{eq:enzymes}].

For a 1d system, the thermodynamic coarsening process described by the CH model is extraordinarily slow with the average droplet radius growing only logarithmically with time~\cite{Bray1994}. 
Hence, one expects that (even weak) enzymatic processes can interrupt coarsening. 
Indeed, solving the dynamics of multiple droplets analytically in the adiabatic limit~\cite{supplement}, we find enzyme fluxes between pairs of differently sized droplets, which are proportional to the difference in the substrate concentration at their closest interfaces.
These currents stop the coarsening process and lead to a steady state where the droplets position themselves equidistantly to each other to even out concentration imbalances between all interfaces [Video~4].

\begin{figure}[btp]
\includegraphics{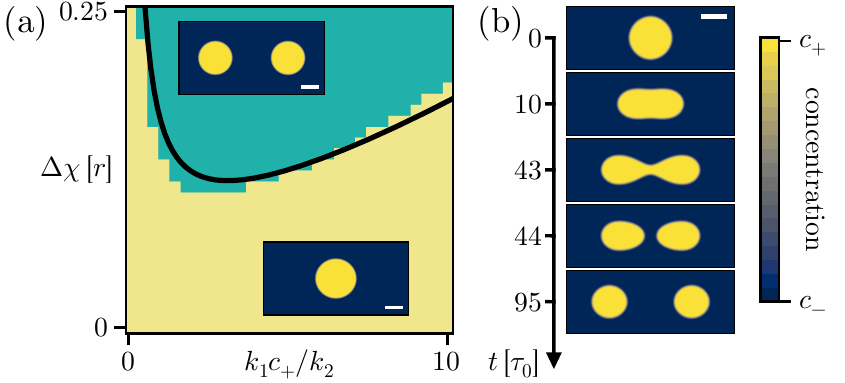}
\caption{\label{fig:coexistence} 
Coexistence and division of 3d droplets.
We consider ${M \epsilon_0 / D = 10}$ and ${w = 0.05 l_0}$. 
Scale bars indicate unit length $l_0 \coloneqq \sqrt{D/k_2}$.
(a) Simulated pairs of droplets with different initial radii $R_1 = 1.1 l_0$ and $R_2 = 0.9 l_0$ either stay separated (cyan regime) or coalesce (yellow regime) depending on ${k_1}$ and on the difference in FH parameters ${\Delta \chi = \chi_p - \chi_s}$.
Solid black line corresponds to analytical estimate, Eq.~\eqref{eq:critical_interaction_parameter}, for $\Delta \chi^* (k_1)$.
(b) Simulation snapshots demonstrating a droplet division in 3d.
We observed droplet divisions only for very strong attraction of enzymes towards substrates, here ${\chi_s/r=-0.5}$. 
}
\end{figure}

For 2d and 3d systems, Ostwald ripening is dominated by surface tension effects (Laplace pressure) and the ensuing law for droplet growth becomes a power law~\cite{Bray1994}.
In this case, one intuitively expects that the coarsening process can be interrupted only if the mass fluxes of the enzymes are sufficiently strongly coupled to the concentration of the products and substrates~\cite{Brauns2021, Weyer2022}. 
Figure~\ref{fig:coexistence}a shows FEM simulation results for a 3d system with a pair of droplets, which confirm this intuitive argument.
The existence of a coarsening threshold for 3d systems can also be understood analytically as a balance between a coarsening current due to surface tension and a mass flux of enzymes driven by reaction-maintained product and substrate concentration gradients.
We estimate the former using the standard Gibbs-Thomson relation~\cite{Bray1994, Weber2019} and the latter by adapting the above results for the 1d system~\cite{supplement}.
By comparing the two currents we find the following estimate for the critical difference between the FH parameters ${\Delta \chi = \chi_p - \chi_s}$ above which one expects droplet coexistence, i.e., interrupted coarsening: 
\begin{equation}
\label{eq:critical_interaction_parameter}
    \Delta \chi^* 
    = 
    \frac{2}{3}
    \frac{
     r w \Delta c / \Delta s^\star
    \left[ l_+^{-1} \cosh(\xi) + l_-^{-1}  \sinh(\xi) \right]^2}
    {l_+^{-1}  
    [\sinh(2 \xi) \,{-}\, 2 \xi] 
    + 
    2 l_-^{-1}  
    [\sinh^2(\xi) \,{-}\, \xi^2] 
    } 
    \, ,
\end{equation}
where ${\xi \coloneqq \bar R/l_+}$ is the ratio of the average droplet radius to the typical length scale of the concentration gradients inside of a droplet, and $\Delta s^\star$ is the difference between the local equilibria of substrate in the two phases~\cite{supplement}.
Notwithstanding the partially heuristic nature of the derivation, our estimate yields a good approximation for the boundary between coexistence and coarsening in parameter space [Fig.~\ref{fig:coexistence}a].

In our numerical simulations, we also observed that initially spherical droplets can undergo a shape instability and elongate in one direction for large enough values of $\Delta \chi$~\cite{supplement}.
Once sufficiently elongated, 3d droplets form a neck and divide [Fig.~\ref{fig:coexistence}, Video 5], which we speculate to occur through a pearling instability driven by surface tension~\cite{Rayleigh1878} independent of the preceding shape instability. 
This droplet division process is driven by intermolecular interactions that induce conservative enzyme fluxes, as opposed to Ref.~\cite{Zwicker2016} where the droplet material is cyclically produced and degraded, leading to non-conservative fluxes and droplet growth.

We have analyzed the nonequilibrium dynamics of enzyme-enriched condensates, whose enzymatic activity guides the generation of inhomogeneous substrate and product concentration profiles that, in turn drive condensate motion.
Conceptually, this corresponds to a feedback mechanism in which, for example, an NTPase such as PomZ undergoing a cycle of hydrolysis (${s \rightarrow p}$, catalyzed by a NAP $c$) and nucleotide exchange (${p \rightarrow s}$) generates concentration gradients of its two different chemical states ($s$ and $p$) that drive droplet movement through a process akin to chemophoresis.
Our results show that such a generic mechanism results in equidistant positioning of condensates in closed domains, persistent condensate motion, and even shape instabilities that lead to condensate division.
We speculate that this mechanism, in its basic form, may be relevant for processes like midcell localization of protein clusters in some prokaryotic cells~\cite{Schumacher2017, Bergeler2018}, directed motion of partition complexes~\cite{Hanauer2021}, equidistant placing of plasmids along nucleoids~\cite{Ietswaart2014}, and maybe even transcription regulation~\cite{HenningerShrinivas2021}.\\


\begin{acknowledgments}
We thank Dominik Schumacher and Lotte S\o{}gaard-Andersen for helpful discussions. 
We acknowledge financial support by the German Research Foundation (DFG) through TRR 174 (Project ID No.~269423233) and SFB1032 (Project ID No.~201269156) and the Excellence Cluster ORIGINS under Germany’s Excellence Strategy (EXC-2094-390783311). 
AG was supported by a DFG fellowship through the Graduate School of Quantitative Biosciences Munich (QBM). 
During his time at the Massachusetts Institute of Technology, AG was supported by the National Science Foundation (NSF) through grant number 2044895.
IM has received funding from the European Union's Framework Programme for Research and Innovation Horizon 2020 under the Marie Sk\l{}odowska-Curie Grant Agreement No. 754388 (LMU Research Fellows) and from LMU excellent, funded by the Federal Ministry of Education and Research (BMBF) and the Free State of Bavaria under the Excellence Strategy of the German Federal Government and the L\"ander.
\end{acknowledgments}

\bibliography{droplets}

\end{document}



\title{Enzyme-enriched condensates show self-propulsion, positioning, and coexistence}
\author{Leonardo Demarchi}
\thanks{These authors contributed equally to this work.}
\altaffiliation[Present address: ]{Sorbonne Université, CNRS, Institut de Biologie Paris-Seine (IBPS), Laboratoire Jean Perrin (LJP), F-75005 Paris, France}
\affiliation{Arnold Sommerfeld Center for Theoretical Physics and Center for NanoScience, Department of Physics, Ludwig-Maximilians-Universit\"at M\"unchen, Theresienstra\ss e 37, D-80333 M\"unchen, Germany}
\author{Andriy Goychuk}
\thanks{These authors contributed equally to this work.}
\altaffiliation[Present address: ]{Institute for Medical Engineering and Science, Massachusetts Institute of Technology, Cambridge, MA 02139, United States}
\email{andriy.goychuk@gmail.com}
\affiliation{Arnold Sommerfeld Center for Theoretical Physics and Center for NanoScience, Department of Physics, Ludwig-Maximilians-Universit\"at M\"unchen, Theresienstra\ss e 37, D-80333 M\"unchen, Germany}
\author{Ivan Maryshev}
\affiliation{Arnold Sommerfeld Center for Theoretical Physics and Center for NanoScience, Department of Physics, Ludwig-Maximilians-Universit\"at M\"unchen, Theresienstra\ss e 37, D-80333 M\"unchen, Germany}
\author{Erwin Frey}
\email{frey@lmu.de}
\affiliation{Arnold Sommerfeld Center for Theoretical Physics and Center for NanoScience, Department of Physics, Ludwig-Maximilians-Universit\"at M\"unchen, Theresienstra\ss e 37, D-80333 M\"unchen, Germany}
\affiliation{Max Planck School Matter to Life, Hofgartenstraße 8, D-80539 M\"unchen, Germany}


\maketitle

\tableofcontents

\clearpage\newpage


\section{Nonequilibrium enzyme dynamics coupled to product and substrate}
\begin{figure}[t]
\centering
\includegraphics{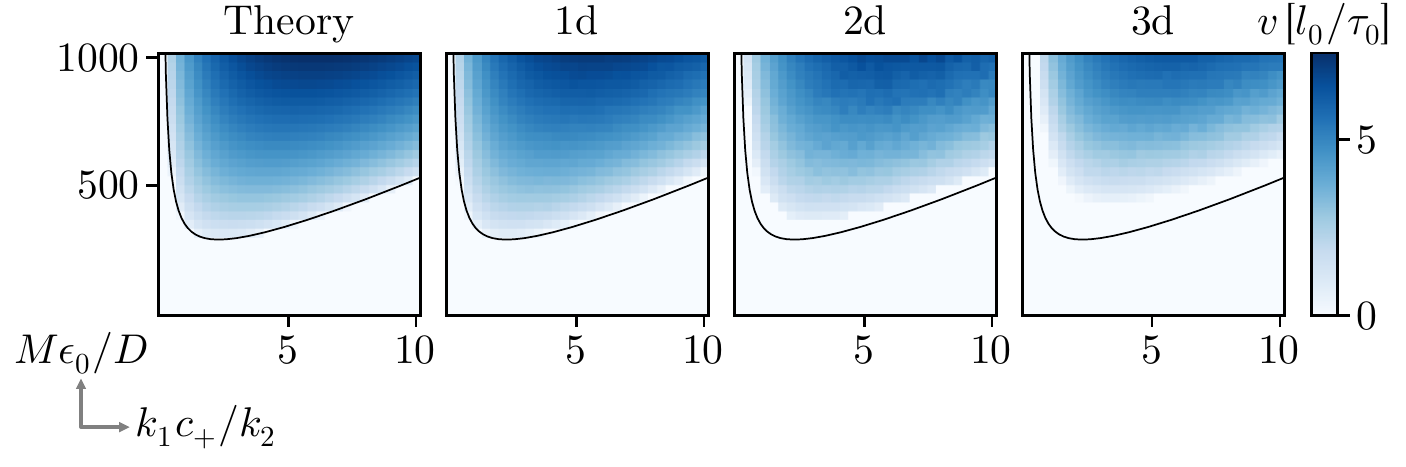}
\caption{%
Theoretical prediction and simulation results for the self-propulsion velocity (color scale) with $M$ and $k_1$ as free parameters. 
The solid black lines indicate the critical mobility $M^*$.
We consider ${s+p = 0.01 c_+}$, ${\chi_s = -5 r}$ and ${\chi_p = -r}$, so that the concentrations of substrate and product are comparatively small.
These results are analogous to Fig.~2c in the main text.
}
\label{fig:self_propulsion_small_concentrations}
\end{figure}

We consider a scenario where there is an interplay between equilibrium phase separation and nonequilibrium chemical reactions.
To model the aspect of equilibrium phase separation, we construct a free energy functional that describes a regular solution consisting of solvent, enzymes, substrates, and products.
For simplicity, we choose the concentrations of substrates,  $s(\boldsymbol{x},t)$, and products, $p(\boldsymbol{x},t)$, to be very small, so that they each contribute to the entropy independent of other concentrations.
Then, gradients in the concentration of substrates and products are also small, so that we can neglect the corresponding interfacial energy terms.
Furthermore, we assume that the interactions between solvent, substrates, and products are negligible.
Under these assumptions, neither the substrate nor the product molecules can show phase separation.
In contrast, we expand the free energy functional that describes the entropy and interactions between solvent and enzymes%
\footnote{We have chosen to perform simulations with comparable values of the three concentration fields, but one can rescale $s$ and $p$ without changing the dynamics, by rescaling $\chi_{s,p}$ and $\Lambda$ accordingly.  Fig.~\ref{fig:self_propulsion_small_concentrations} shows the results of simulations with rescaled parameters.}, which are present at much larger concentrations than substrates and products and are assumed to exhibit phase separation, up to fourth order in the enzyme concentration, $c(\boldsymbol{x},t)$, and use the Ginzburg-Landau functional
\begin{equation}
\label{eq:freeenergy_cahnhilliard}
    f_0(c) = \frac{u}{4}(c-\tilde c)^4 - \frac{r}{2}(c-\tilde c)^2 + \frac{\kappa}{2} |\boldsymbol\nabla c|^2 \, ,
\end{equation}
with a positive stiffness, $\kappa > 0$, and parameters $u>0$ and $r>0$ (corresponding to a double-well potential).
Finally, we parameterize enzyme--substrate and enzyme--product interactions with the Flory-Huggins (FH) parameters $\chi_s$ and $\chi_p$, respectively. 
In summary, the statistics of the system is characterized by the following effective free energy functional,
\begin{equation} 
\label{eq:freeenergy}
    \mathcal{F} 
    = 
    \int\! d^d\boldsymbol{x} \, 
    \Big\{ 
    k_B T \bigl[ s \log(s \, \nu) + p \log(p \, \nu) \bigr] + \chi_s \, c \, s + \chi_p \, c \, p + f_0(c) 
    \Big\} 
    \, ,
\end{equation}
where $k_B T$ is the thermal energy and $d$ refers to the number of spatial dimensions.
For simplicity, we have here assumed that substrates and products have the same molecular volume $\nu$.

Given the above constraints on the model parameters, the enzymes will show spontaneous phase separation into droplets or labyrinth-like patterns if the average enzyme concentration lies in the spinodal regime, or can separate through nucleation and growth in the binodal regime.
For our simulations, we initialize the system with a single predefined droplet to study its dynamics, or with multiple droplets to study their coarsening dynamics.

An exchange of particles (substrates, products, or enzymes) modifies the free energy of the system and is thus associated with the following chemical potentials:
\begin{subequations}
\label{eq:chemical_potentials}
\begin{align}
    \mu_s &= \frac{\delta\mathcal{F}}{\delta s} = k_B T \bigl[\log(s \, \nu) + \nu^{-1}\bigr] + \chi_s \, c 
    \, , \\
    \mu_p &= \frac{\delta\mathcal{F}}{\delta p} = k_B T \bigl[\log(p \, \nu) + \nu^{-1}\bigr] + \chi_p \, c \, , \\
    \mu_c &= \frac{\delta\mathcal{F}}{\delta c} = \mu_0(c) + \chi_s \, s + \chi_p \, p \, ,
\end{align}
\end{subequations}
where we have defined the chemical potential of the Cahn-Hilliard model, 
\begin{equation}
    \mu_0(c) \coloneqq u \, (c-\tilde c)^3 - r \, (c-\tilde c) - \kappa \, \nabla^2 c
\end{equation}
to simplify our notation.
For our analysis, we consider a scenario where enzymes interact with substrates and products weakly, ${|\chi_{s} s| + |\chi_{p} p| \ll r \, (c_+ - c_-)}$, where $c_\pm$ refers to the concentrations in the high-density and in the low-density phase, respectively.
Then, the chemical potential $\mu_0$ dominates the phase separation of enzymes and maintains their sharp concentration profile.
Following the general ideas of nonequilibrium thermodynamics~\cite{DeGroot2013, Balian2006}, gradients in the chemical potentials~\eqref{eq:chemical_potentials} drive conservative currents that gradually minimize the free energy functional $\mathcal{F}$,
\begin{equation} 
\label{eq:constitutive}
\begin{bmatrix}
    \boldsymbol{j}_s \\
    \boldsymbol{j}_p \\
    \boldsymbol{j}_c
\end{bmatrix} 
=
-
\begin{bmatrix}
    \Lambda \, s & 0 & 0 \\
    0 & \Lambda \, p & 0 \\
    0 & 0 & M c
\end{bmatrix} 
\cdot
\begin{bmatrix}
    \boldsymbol\nabla \mu_s \\ 
    \boldsymbol\nabla \mu_p \\
    \boldsymbol\nabla \mu_c
\end{bmatrix} \, ,
\end{equation}
where substrates and products are for simplicity assumed to have identical mobility $\Lambda$, and enzymes have mobility $M$ (Onsager coefficients).
Equation~\eqref{eq:constitutive} can be interpreted as the local concentration of each species multiplied with an average drift velocity induced by driving forces $\boldsymbol\nabla \mu$, so that the diagonal entries in the response matrix are the quotient of the local concentration and the viscous friction of each species.

We consider a system that is driven out of equilibrium by chemical reactions that are not derived from the free energy functional $\mathcal{F}$ and therefore provide an external energy influx.
In particular, we consider a scenario where enzymes catalyze turnover of substrates into products with rate $k_1  c \, s$, while products gradually decay into substrates with rate $k_2 \, p$.
Then, the dynamics of substrates, products, and enzymes is given by:
\begin{equation}
    \partial_t s + \boldsymbol\nabla \cdot \boldsymbol{j}_s = k_2 \, p - k_1 \, c \, s \,, \quad
    \partial_t p + \boldsymbol\nabla \cdot \boldsymbol{j}_p = k_1 \, c \, s - k_2 \, p \,,
    \quad \text{and} \quad
    \partial_t c + \boldsymbol\nabla \cdot \boldsymbol{j}_c = 0 \, .
\end{equation}
Taking everything together, one arrives at Eqs.~(2)~and~(3) of the main text:
\begin{subequations}
\label{eq:system}
\begin{align}
\label{eq:enzymes}
    \partial_t c &= \boldsymbol\nabla \cdot \Bigl\{ M c \boldsymbol\nabla \bigl[\mu_0(c) + \chi_s \, s + \chi_p \, p \bigr]\Bigr\} \,, \\
\label{eq:substrates}
    \partial_t s &= D \nabla^2 s - k_1 c s + k_2 p + \Lambda \chi_s \boldsymbol\nabla \cdot (s \boldsymbol\nabla c) \,, \\
\label{eq:products}
    \partial_t p &= D \nabla^2 p + k_1 c s - k_2 p + \Lambda \chi_p \boldsymbol\nabla \cdot (p \boldsymbol \nabla c) \,,
\end{align}
\end{subequations}
where we have related the mobility of substrates and products to their diffusion coefficient by the Einstein relation, $D=\Lambda k_B T$.
In the following, we take the liberty of formally decoupling the diffusion coefficient $D$ from the mobility $\Lambda$.
In doing so, we introduce a further source of far from equilibrium dynamics by breaking the fluctuation-dissipation relation valid for thermal equilibrium systems.

\section{Numerical simulations of single droplet dynamics}
\label{sec::simulations}
We solved the system of partial differential equations~\eqref{eq:system} numerically; the code is available on~\cite{CodeGithub}.
To that end, we used the implicit Euler method for time discretization and the finite element method for spatial discretization; for the latter, we used the FEniCS libraries~\cite{Logg2012}.
We considered a finite-sized domain with no-flux boundary conditions, which in one dimension is parameterized by $x \in [-L, L]$.
As initial conditions, we chose uniform concentration profiles for the substrates and products, and controlled the average concentration of these two species $s(x) + p(x) = n$.
Furthermore, we initialized the enzyme concentration profile to resemble a single droplet with a sharp interface:
\begin{equation}
\label{eq:enzyme_concentration_sharp}%
c_\text{sharp}(x) =
\begin{cases} 
    c_+ \,, &\text{for } x \in [-R, R] 
    \,, \text{and} \\
    c_- \,, &\text{otherwise}.
\end{cases}
\end{equation}
As indicated in the main text, we use $c_+$ as reference concentration and define the characteristic time $\tau_0 \coloneqq k_2^{-1}$, diffusion length in the absence of enzymes $l_0 \coloneqq \sqrt{D/k_2}$, and reference energy $\epsilon_0 \coloneqq r c_+$.
The remaining parameters, ${c_- = 0.1 c_+}$, ${w = 0.1 l_0}$, ${R = l_0}$, ${L = 5 l_0}$, ${M = 100 D/\epsilon_0}$, ${k_1 = k_2/c_+}$, ${\chi_s = -0.05 r}$, ${\chi_p = -0.01 r}$, and ${s+p = c_+}$, are fixed unless stated otherwise.
We then simulated the system for a total time of $100 \, \tau_0$ to let the system reach a steady state.
Figure~1 in the main text shows that the enzymes maintain a single droplet with a steep interface, enriching products at the expense of substrates in regions with high enzyme concentration.
Outside of the droplet, where the enzyme concentration is low, substrates are replenished at the expense of products.

\subsection{Self-propulsion of a droplet}
Our simulations demonstrate that droplets can self-propel and sustain motion for a wide range of parameters, in 1d [Fig.~2 in the main text, Supplemental Video~\ref{vid:self_propulsion}] as well as 2d and 3d [Supplemental Video~\ref{vid:self_propulsion}].
We initialized droplets at the origin of an interval of half-length $L=10 \, l_0$ in 1d and at the center of a circular domain of radius $L_r = 7 \, l_0$ in 2d.
To speed up simulations in 3d, we chose a rotationally invariant cylindrical coordinate system with radius $L_r = 4 \, l_0$ and height $L_z = 7 \, l_0$, which reduced the number of degrees of freedom but also constrained the space of permitted solutions.
In all of our simulations, breaking the symmetry of the droplet requires an initial perturbation.
We provided such a perturbation through spatial inhomogeneities in the concentration profiles of substrates and products, which are small compared to the average total concentration $n \coloneqq \left\langle s(\boldsymbol{x}) + p(\boldsymbol{x}) \right\rangle_{\boldsymbol{x}}$.

\subsection{Self-centering of a droplet in a finite domain}
\label{sec:self_centering_simulations}
In a finite domain, our simulations show that a self-propelling droplet can either reach and adhere to one of the domain boundaries, or exhibit oscillatory motion by reorienting at the domain boundaries [Fig.~3a in the main text].
Reorientation at the domain boundaries is mediated by an effective repulsion, which has the following mechanistic origin.
In our simulations, we observe that the enzymatic depletion of substrate is enhanced in the vicinity of a domain boundary, where substrate resupply through diffusive currents is limited.
The resulting substrate concentration gradients induce net enzyme currents, through attractive substrate--enzyme interactions, away from the domain boundary.

Next, we study how a droplet that does not meet the parameter criteria for self-propulsion [discussed in detail in section~\fullref{sec::self_consistency_nonreciprocal}] will position itself in the domain.
To that end, we considered a droplet whose center $x_d(0)=-l_0$ is initially offset from the domain center.
We initialized the distribution of substrates and products in the steady state that is reached in the absence of interactions, $\chi_{s,p} = 0$, by performing ``pre-simulations'' for a duration of $1000 \, \tau_0$.
Then, we introduced interactions $\chi_{s,p} \neq 0$ and studied the resulting trajectory of the droplet center in a one-dimensional domain as a function of time, $x_d(t)$ [Fig.~3a in the main text, 1d droplet dynamics shown in Supplemental Video~\ref{vid:positioning_1d} while 2d and 3d droplet dynamics shown in Supplemental Video~\ref{vid:positioning_2d_3d}]. 
To improve the performance of the simulations, we used adaptive time stepping, and confirmed that the total simulation time is sufficiently long for the droplets to either attach to the boundary, perform several oscillations, or relax to the domain center.
For droplets that do not exhibit self-propulsion, instead of sustained oscillations we observed gradual localization to the domain center akin to a damped harmonic oscillator.
To account for both self-sustained and damped oscillatory motion, we fitted each droplet trajectory with an exponentially damped sinusoidal curve using the library LMFIT~\cite{Newville2014}: $x_d(t) = A \, e^{-\lambda t} \, \cos(\omega t + \phi)$, where $A$ is the amplitude, $\lambda$ the decay rate, $\omega$ the frequency and $\phi$ the initial value of the phase.
Trajectories where the droplet attaches to the domain boundary cannot be fitted in such a way, and were therefore excluded.
The resulting estimates for the decay rate and for the frequency of the oscillations are shown in Fig.~3b in the main text as functions of the droplet mobility $M$.

\begin{figure}[t]
\centering
\includegraphics{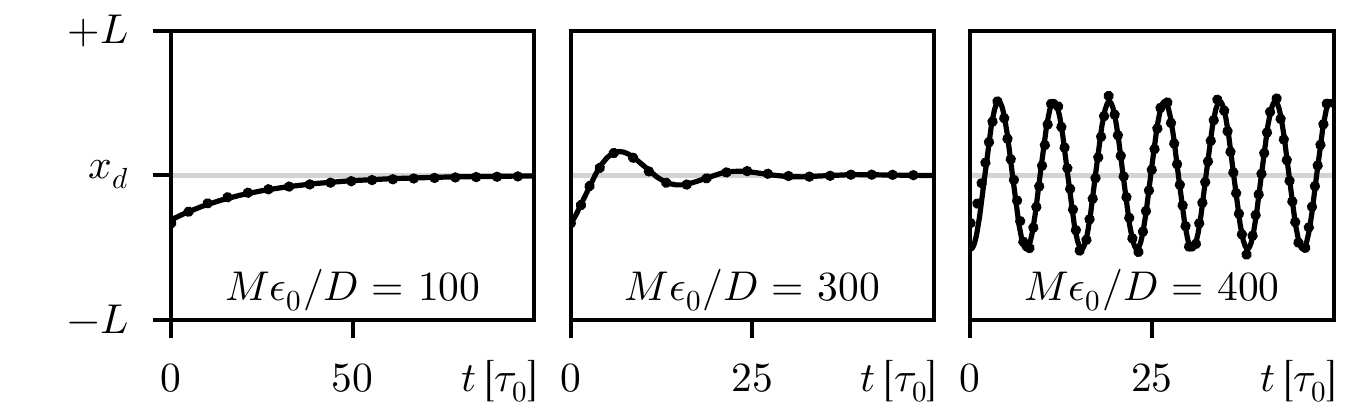}
\caption{%
Comparison of the droplet trajectories from the simulations (dots) with the best fitting curves (solid lines), for the panels shown in Fig.~3a in the main text.
}
\label{fig:trajectories_fit_comparison}
\end{figure}

\section{Analytic theory of a single droplet in a one-dimensional domain}
\label{sec:analytic_theory_droplet_1d}
For our theoretical analysis in the present paper, we restrict ourselves to a 1d system.
A 2d and 3d analysis is also possible with a semi-analytical approach, but exceeds the scope of the present paper and will be published elsewhere~\cite{DropletsLongPaper}.
The idea of our theoretical analysis rests on two pillars:
\begin{enumerate*}[label=(\roman*)]
    \item In all our simulations, we observed that the droplet maintains a steep interface.
    This suggests that one can well describe our system analytically with a sharp-interface approximation.
    \item Similar to the analysis of Fisher waves~\cite{FISHER1937}, we treat moving droplets through a transformation into the co-moving frame.
\end{enumerate*}
As we discuss next, these considerations allow us to derive the concentration profiles of substrates and products, as well as a self-consistency relation for the droplet velocity $v$.

\subsection{Substrate and product redistribution by a stationary droplet}
\label{sec::sharp_interface}
To perform our theoretical analysis, we first determine the steady-state concentration profiles of substrates and products in response to the presence of an enzymatic droplet; see Eqs.~\eqref{eq:substrates}~and~\eqref{eq:products}.
Because the enzyme concentration profile is well described by a sharp-interface approximation, Eq.~\eqref{eq:enzyme_concentration_sharp}, for now we do not need to explicitly study the dynamics of the enzymes, Eq.~\eqref{eq:enzymes}.

In the scenario of nonreciprocal interactions with $\Lambda = 0$, by summing Eqs.~\eqref{eq:substrates}~and~\eqref{eq:products} one finds that the total density of substrates and products is uniform in space, $s(x,t) + p(x,t) = n$; the general scenario with $\Lambda \neq 0$ will be analyzed elsewhere~\cite{DropletsLongPaper}.
By substituting this conservation law into Eq.~\eqref{eq:substrates}, we find that the steady-state distribution of substrates is determined by
\begin{equation}
\label{eq:stationarysubstrate}
D \, \partial_x^2 s(x) - \bigl[k_1 c_\text{sharp}(x) + k_2\bigr] s(x) + k_2 n = 0 \, ,
\end{equation}
where $c_\text{sharp}(x)$ refers to the concentration profile of enzymes in the sharp-interface approximation [Eq.~\eqref{eq:enzyme_concentration_sharp}].
The resulting steady-state distribution of substrates also defines the concentration profile of products $p(x,t) = n - s(x,t)$.

\begin{figure}[t]
\centering
\includegraphics{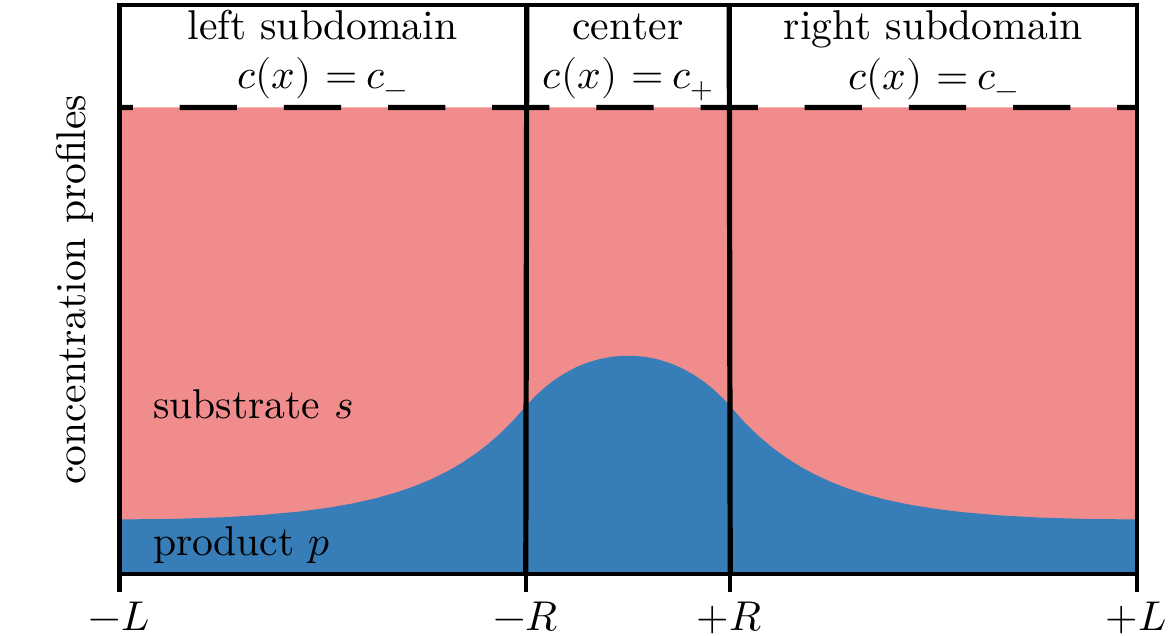}
\caption{%
Illustration of the three sub-domains in the sharp-interface approximation $c(x) = c_\text{sharp}(x)$.
The total concentration of substrates and products is conserved, $s(x,t) + p(x,t) = n$.
In each subdomain, we then only need to solve a Helmholtz equation.
}
\label{fig:illustration_helmholtz}
\end{figure}

Because the enzyme concentration is piecewise constant in the sharp-interface approximation, Eq.~\eqref{eq:stationarysubstrate} reduces to a Helmholtz equation defined on three subdomains [Fig.~\ref{fig:illustration_helmholtz}].
The concentration profiles of substrates and products then have different characteristic diffusion lengths, ${l_+ \coloneqq \sqrt{D/(k_1 c_+ + k_2)}}$ inside (center) and ${l_- \coloneqq \sqrt{D/(k_1 c_- + k_2)}}$ outside (left and right) of the droplet.
Solving the corresponding Helmholtz equation, Eq.~\eqref{eq:stationarysubstrate}, the distribution of substrates for a stationary droplet is given by
\begin{equation}
\label{eq:substrate_solution}
s(x) = 
\begin{dcases}
    \frac{k_2 \, n}{D} \, l_+^2 + 2 \, A_\text{in} \, \cosh\left[\frac{x}{l_+}\right] \,, &\text{for } \|x\| \leq R\,, \\
    \frac{k_2 \, n}{D} \, l_-^2 + A_\text{out} \, \exp\left[-\frac{|x|}{l_-}\right] + B_\text{out} \, \exp\left[\frac{|x|}{l_-}\right]  \,, &\text{otherwise},
\end{dcases}
\end{equation}
where $A_\text{in/out}$ and $B_\text{out}$ are integration constants.
We determined these integration constants by imposing smoothness and continuity of the concentration profiles at the droplet interfaces $x = \pm R$, which separate different subdomains, as well as no-flux boundary conditions at the domain boundaries $x = \pm L$.
We provide the full (and rather lengthy) expressions for these integration constants in~\cite{CodeGithub}.
However, even without having these explicit expressions at hand, one can nevertheless analyze generic features of the substrate and product concentration profiles dictated by Eq.~\eqref{eq:substrate_solution}.

Because all concentrations must remain finite in the far field $x \rightarrow \pm \infty$, the integration constant $B_\text{out}$ must vanish ($B_\text{out} = 0$) for an infinitely large domain ($L\rightarrow \infty$); see Eq.~\eqref{eq:substrate_solution}.
Therefore, it follows from Eq.~\eqref{eq:substrate_solution} that the substrate and product concentrations far away from the droplet ($x \rightarrow \pm \infty$) are given by their local reactive equilibria%
\footnote{We define the reactive equilibria as the solution of the reaction-diffusion equations for substrates and products, Eq.~\eqref{eq:substrates}~and~\eqref{eq:products}, in the limit of vanishing transport $D\rightarrow 0$ and $\Lambda \rightarrow 0$.}, $s(\pm\infty) = \frac{k_2 \, n}{D} \, l_-^2 = \frac{k_2 \, n}{k_1 c_- + k_2}$, which correspond to the solution of the Helmholtz equation~\eqref{eq:stationarysubstrate} in the limit $D \rightarrow 0$.
Only the concentrations at the center of the droplet, $s(0) = \frac{k_2 \, n}{D} \, l_+^2 + 2 \, A_\text{in}$, are shifted by $2A_\text{in}$ relative to their local reactive equilibria.
The magnitude of this shift depends on the relative size of the droplet compared to the characteristic length of the concentration profiles, $R / l_+$, and must vanish for very large droplets\footnote{This can be seen by taking the limit of $D \rightarrow 0$ in Eq.~\eqref{eq:stationarysubstrate}.} $R \gg l_+$.
The difference between the reactive equilibria of substrate at low and at high enzyme concentration is given by:
\begin{equation}
\label{eq:reactive_equilibria_difference}
    \Delta s^\star 
    \coloneqq \frac{k_2 \, n}{D} \, \left(l_-^2 - l_+^2\right) 
    = \frac{k_2 \, n}{D} \, \frac{k_1 \, \Delta c \, l_+^2 l_-^2}{D} \, .
\end{equation}
Thus, the level of substrate depletion in the droplet is proportional to its enrichment of enzymes relative to the surrounding solution, $\Delta c = c_+ - c_-$, and analogously implies different characteristic diffusion lengths inside and outside of the droplet, $l_+ \neq l_-$.

Having discussed these features of the substrate and product concentration profiles, we compare our analytic predictions (in the sharp-interface approximation) to our numeric results (with a diffuse interface) and find very good agreement [Fig.~1 in the main text]. 
Because of this excellent agreement, in the following, we use the sharp-interface approximation to also study more intricate scenarios where the droplet becomes mobile and positions itself in the domain.

\subsection{Substrate and product redistribution by a moving droplet}
\label{sec::sharp_interface_moving}
We now generalize the results of the former section to an enzymatic droplet that moves with constant velocity $v$, where the enzyme concentration profile can be written in the form of a travelling wave ${c(x,t) = c_\text{sharp}(x - v t)}$; see Eq.~\eqref{eq:enzyme_concentration_sharp}.
As we have discussed before, the total density of substrates and products is uniform in space, $s(x,t) + p(x,t) = n$, for nonreciprocal interactions $\Lambda = 0$. 
Using the substitution $z = x - v t$ to transform Eq.~\eqref{eq:substrates} into the co-moving reference frame of the droplet, the steady-state distribution of substrates is then determined by
\begin{equation}
    \label{eq:movingsubstrate}
    D \, \partial_z^2 s(z) - \bigl[k_1 c_\text{sharp}(z) + k_2\bigr] s(z) + k_2 n = - v \, \partial_z s(z) \, ,
\end{equation}
which differs from Eq.~\eqref{eq:stationarysubstrate} only by an advection term.
To simplify the expressions in the following, we introduce the Péclet number, $\text{Pe}_0 = v R/D$, and two modified Péclet numbers that include a correction for the characteristic lengths $l_\pm$ of the concentration profiles inside and outside of the droplet, $\text{Pe}_\pm = \sqrt{\text{Pe}_0^2 + (2R / l_\pm)^2}$.
Furthermore, we note that the distribution of substrates must remain finite in the far field $z\rightarrow\pm\infty$.
Then, solving Eq.~\eqref{eq:movingsubstrate} in each of the three subdomains [Fig.~\ref{fig:illustration_helmholtz}], imposing smoothness and continuity at the subdomain interfaces, and assuming an infinitely large overall domain, yields the steady-state distribution of substrates in the co-moving frame of the droplet:
\begin{equation}
\label{eq:substrate_comoving_frame}
s(z) = 
\begin{dcases}
    \frac{k_2 \, n}{D} \, l_+^2 + A_\text{in} \, \exp\left[-\frac{\text{Pe}_0 - \text{Pe}_+}{2} \frac{z}{R} \right] + B_\text{in} \, \exp\left[-\frac{\text{Pe}_0 + \text{Pe}_+}{2} \frac{z}{R}\right] &\text{for } \| z\| \leq R\,, \\
    \frac{k_2 \, n}{D} \, l_-^2 + A_\text{out} \, \exp\left[-\frac{\text{Pe}_0 - \text{Pe}_-}{2} \frac{z}{R}\right]   &\text{for } z < -R\,, \\
    \frac{k_2 \, n}{D} \, l_-^2 + B_\text{out} \, \exp\left[-\frac{\text{Pe}_0 + \text{Pe}_-}{2} \frac{z}{R}\right] \,, &\text{otherwise},
\end{dcases}
\end{equation}
which also defines the concentration profile of the products $p(x,t) = n - s(x,t)$.
As before, we determined the integration constants $A_\text{in/out}$ and $B_\text{in/out}$ by imposing smoothness and continuity at the droplet interfaces $z = \pm R$.
To that end, we used the Python library Sympy~\cite{Meurer2017} for symbolic calculations and confirmed with the computer algebra system Mathematica~\cite{Mathematica}, and provide the full expressions in~\cite{CodeGithub}.

The concentration profiles of substrates and products, in the co-moving frame of an enzymatic droplet with velocity $v$, are plotted in Fig.~2a of the main text.
We observe that the droplet enriches the concentration of products at the expense of substrates, which is most pronounced at the trailing edge of the droplet.
This leads to a difference in substrate concentrations between the right and the left edge of the droplet, $\Delta s(v) \coloneqq s(R) - s(-R)$, which we quantify after inserting the integration constants~\cite{CodeGithub} in Eq.~\eqref{eq:substrate_comoving_frame}:
\begin{equation}
\label{eq:deltacs_lambda0}
    \Delta s(v) = \Delta s^\star \, \frac{\text{Pe}_+ \text{Pe}_0 \, \bigl[\cosh(\text{Pe}_+) - \cosh(\text{Pe}_0) \bigr] + \text{Pe}_- \bigl[ \text{Pe}_0 \sinh(\text{Pe}_+) - \text{Pe}_+ \sinh(\text{Pe}_0) \bigr]}{\text{Pe}_+ \text{Pe}_- \cosh(\text{Pe}_+) + (\text{Pe}_+^2 + \text{Pe}_-^2) \sinh(\text{Pe}_+) / 2 } \, .
\end{equation}
Here, the substrate concentration difference between the reactive equilibria, $\Delta s^\star$, is given by Eq.~\eqref{eq:reactive_equilibria_difference}.
Because the total density of substrates and products is uniform in the scenario with nonreciprocal interactions ($\Lambda = 0$), the difference in product concentrations between the right and the left edge of the droplet is given by $\Delta p(v) = - \Delta s(v)$.
As we show next, these concentration differences can drive a net flux of enzymes from the trailing edge towards the leading edge of the moving droplet, thus sustaining its motion.
The more generic scenario with $\Lambda \neq 0$ exceeds the scope of the present paper and will be published elsewhere~\cite{DropletsLongPaper}.
In short, one finds that $\Lambda \neq 0$ can place further constraints on the parameters to observe self-propulsion.

\subsection{Gradients in substrate and product concentration drive droplet motion}
\label{sec::induced_flux}
In the discussion so far, we have used the sharp-interface approximation for the enzyme concentration profile $c(x,t) = c_\text{sharp}(x - v t)$ of a droplet that moves with constant velocity $v$; see Eq.~\eqref{eq:enzyme_concentration_sharp}.
Now, we investigate the premises for this sharp-interface approximation to be consistent with the enzyme dynamics obeying a continuity equation~\eqref{eq:enzymes}.
The sharp-interface approximation implies that the moving concentration profile of enzymes must be in steady state.
Using the substitution $z = x - v t$ to transform the continuity equation~\eqref{eq:enzymes} into the co-moving reference frame of the droplet, one then has:
\begin{equation}
\label{eq:stationaryenzyme}
     \partial_z \Bigl\{ M c(z) \, \partial_z \bigl[\mu_0 (c(z)) + \chi_s \, s(z) + \chi_p \, p(z)\bigr] \Bigr\} = - v \, \partial_z c(z) \equiv - \partial_z j(z) \, .
\end{equation}
Indefinite integration of Eq.~\eqref{eq:stationaryenzyme} yields
\begin{equation}
\label{eq:stationaryenzyme_integrated}
     M c(z) \, \partial_z \bigl[\mu_0 (c(z)) + \chi_s \, s(z) + \chi_p \, p(z)\bigr] = - v \, c(z) + j_0 \, ,
\end{equation}
up to an integration constant $j_0$.
To determine this integration constant, we first note that far away from the droplet, at $z\pm\infty$, all concentration profiles are homogeneous and the enzyme concentration is given by $c(\pm\infty) = c_-$.
Comparing these far-field conditions with Eq.~\eqref{eq:stationaryenzyme_integrated}, one finds that $j_0 = v \, c_-$.
To summarize, we have so far
\begin{equation}
\label{eq:stationaryenzyme_fluxes}
     M c(z) \, \partial_z \bigl[\mu_0 (c(z)) + \chi_s \, s(z) + \chi_p \, p(z)\bigr] = - v \, \bigl[ c(z) - c_- \bigr] \, .
\end{equation}
Now, we use the sharp-interface approximation $c(z) = c_\text{sharp}(z)$; see Eq.~\eqref{eq:enzyme_concentration_sharp}.
Integrating Eq.~\eqref{eq:stationaryenzyme_fluxes} over the droplet $z\in [-R, R]$ to obtain an expression independent of \emph{local} gradients, we find
\begin{equation}
    v = -\frac{M c_+}{2 R \, \Delta c} 
    \Big[
    \chi_s \, \Delta s(v) + \chi_p \, \Delta p(v)
    \Big] \, ,
    \label{eq:self_consistency}
\end{equation}
where we have defined the enzyme concentration difference $\Delta c \coloneqq c_+ - c_-$.
Note that the bare chemical potential of the enzyme, $\mu_0(z)$, dropped out because it is mirror symmetric in the sharp-interface approximation, so that $\mu_0(R) = \mu_0(-R)$.
Thus, the bare chemical potential $\mu_0(z)$ cannot drive directed net fluxes of enzymes on its own: it instead drives relaxation towards an equilibrium state.
In contrast, the coupling to substrate and enzymes, which are driven out of equilibrium by chemical reactions, drives relaxation towards a non-equilibrium steady state.
Our result summarized by Eq.~\eqref{eq:self_consistency} quantifies how asymmetric substrate and product concentration profiles (see Eq.~\eqref{eq:deltacs_lambda0}) drive droplet motion.

\subsection{A self-consistency relation for the self-propulsion instability}
\label{sec::self_consistency_nonreciprocal}
In the scenario of nonreciprocal interactions ($\Lambda = 0$) where the total density of substrates and products is constant, $s(z) + p(z) = n$, Eq.~\eqref{eq:self_consistency} simplifies to:
\begin{equation}
    v = \frac{M c_+ \, \Delta\chi}{2 R \, \Delta c} \,
    \Delta s(v) \, ,
    \label{eq:self_consistency_nonreciprocal}
\end{equation}
where $\Delta\chi \coloneqq \chi_p - \chi_s$ is a measure for how enzymes are pulled more towards substrates than towards products.
The droplet velocity is proportional to the overall mobility $M$ of enzymes, and for constant $\Delta s$ decreases with the overall number of enzymes $2R \Delta c$ that are translocated.

The droplet velocity in response to a gradient in the substrate concentration, Eq.~\eqref{eq:self_consistency_nonreciprocal}, and the distribution of substrates in response to a droplet moving with a fixed velocity, Eq.~\eqref{eq:deltacs_lambda0}, together form a self-consistency relation.
This self-consistency relation can be solved graphically by identifying a velocity where the left-hand side and the right-hand side of Eq.~\eqref{eq:self_consistency_nonreciprocal} intersect, as illustrated in Fig.~2b in the main text.
We note that the substrate concentration difference between the right and the left edge of the droplet, $\Delta s(v)$, is point symmetric with respect to a reversal of the velocity, $\Delta s(-v)=-\Delta s(v)$, which is a feature of the intrinsic symmetry under parity of the system%
\footnote{One can see this symmetry by making the parity transformations $z\rightarrow -z$ and $v \rightarrow -v$ in Eq.~\eqref{eq:movingsubstrate}, which preserves the structure of the equation and corresponds to a mirrored concentration field $s(-z)$. One could also argue more heuristically: because there is no global bias for breaking symmetry (polarizing) towards the left or to the right, the droplet could go either way with equal velocity.}.
Because of this symmetry, it is sufficient to discuss a scenario with positive velocities only, $v \geq 0$.
In the limit of very large velocities $v\rightarrow \infty$, the substrate and product concentration fields do not have sufficient time to respond to the moving enzymatic droplet and therefore remain homogeneous, which implies $\Delta s(\infty) \rightarrow 0$.

\begin{figure}[t]
\centering
\includegraphics{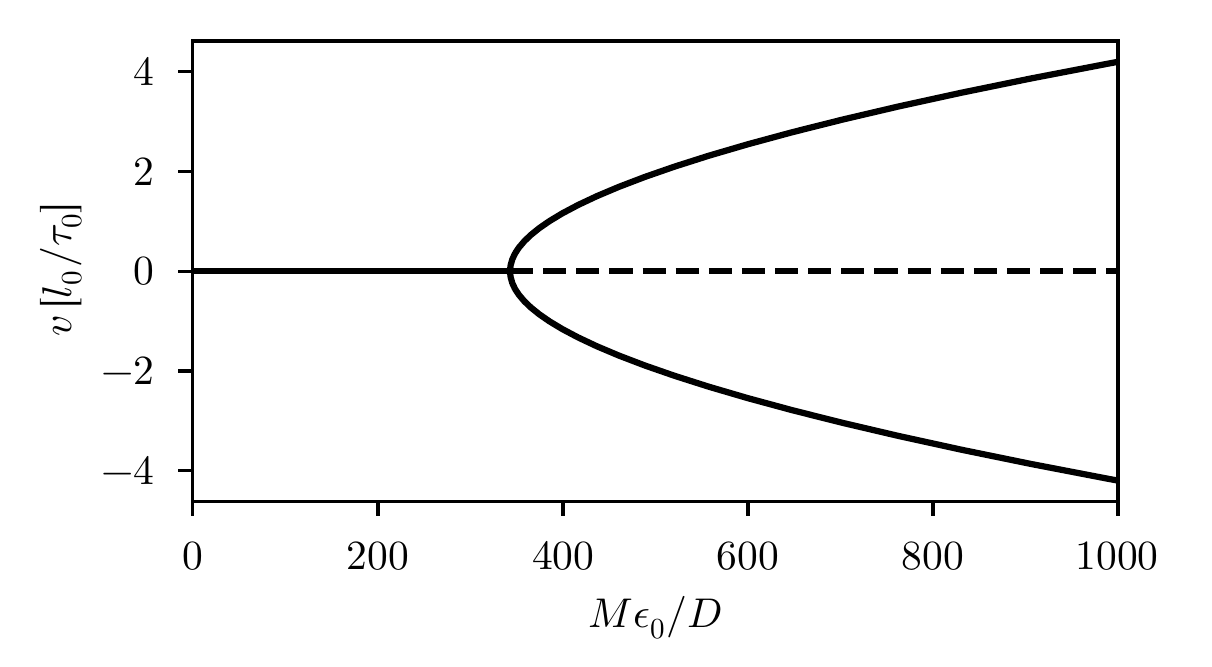}
\caption{%
Pitchfork bifurcation for droplet self-propulsion.
Tuning enzyme mobility $M$ as control parameter, one finds that small enzyme mobilities admit only a single stable solution (solid lines) with vanishing droplet velocity.
This solution becomes unstable above a critical value of the enzyme mobility $M^\star$ (dashed line), where two new stable solutions with finite velocity appear.
}
\label{fig:bifurcation_diagram}
\end{figure}

Because of the point symmetry of $\Delta s(v)$, the self-consistency relation~\eqref{eq:self_consistency_nonreciprocal} always permits a trivial solution with vanishing velocity $v=0$.
For small velocities, we observe that the substrate concentration difference between the leading and the trailing edge of the droplet, $\Delta s(v)$, initially grows with increasing velocity $v$ until it reaches a single maximum, see Eq.~\eqref{eq:deltacs_lambda0} and Fig.~2b in the main text.
Given these features of the function $\Delta s(v)$, two additional non-trivial solutions with finite droplet velocity $|v| \neq 0$ emerge if the right-hand side of Eq.~\eqref{eq:self_consistency_nonreciprocal} grows faster than the left-hand side in the limit of small velocities, $v\rightarrow 0$:
\begin{equation}
\label{eq:self_consistency_nonreciprocal_simplified}
    1 < \frac{M c_+ \, \Delta\chi}{2 R \, \Delta c} \,
    \partial_v\Delta s(v)\biggr\vert_{v=0} \, .
\end{equation}
If a non-trivial solution to the self-consistency relation exists, then an initial inhomogeneity of the substrate concentration profile will self-reinforce through a positive feedback loop with the motion of the droplet.
Because of this feedback loop, the droplet will settle in a state that corresponds to the finite velocity $v$ admitted by the self-consistency relation~\eqref{eq:self_consistency_nonreciprocal}.
Substituting Eq.~\eqref{eq:deltacs_lambda0} into Eq.~\eqref{eq:self_consistency_nonreciprocal_simplified}, the criterion to observe self-propelled droplets in our simulations can therefore be written as:
\begin{equation}
    \label{eq:motility_condition_full}
    1 < \frac{M c_+ \, \Delta\chi \, l_- l_+}{2 R D} \frac{\Delta s^\star}{\Delta c} \, \frac{l_+ \sinh(2 R/l_+) + l_- \cosh(2 R/l_+) - l_- - 2 R}{(l_-^2 + l_+^2) \sinh(2 R/l_+) + 2 l_+ l_- \cosh(2 R/l_+)} \, .
\end{equation}
The emergence of two new stable fixed points $|v| \neq 0$ as a function of enzyme mobility $M$ as control parameter, concomitant with a destabilization of the trivial fixed point $v = 0$, corresponds to a pitchfork bifurcation; see Fig.~\ref{fig:bifurcation_diagram}.

To gain a better understanding on the conditions that are required to observe droplet self-propulsion, we study inequality~\eqref{eq:motility_condition_full} in the limits of very large or very small droplets:
\begin{equation}
    \label{eq:motility_condition_approximate}
    1 < \frac{M c_+ \, \Delta\chi}{2 D} \frac{\Delta s^\star}{\Delta c} \,
    \frac{l_-}{R}
    \color{gray}\times\color{black}
    \begin{dcases}
    \frac{l_+}{l_- + l_+} & , \,\text{for}\, R \gg l_\pm \,,  \\
    \frac{R^2}{l_+^2} & , \,\text{for}\, R \ll l_\pm\, .
    \end{dcases}
\end{equation}
As a function of the droplet radius $R$, the right-hand side of inequality~\eqref{eq:motility_condition_approximate} is clearly non-monotonic: it grows with increasing droplet size for small droplets, but then decays with increasing droplet size for large droplets.
Thus, the self-propulsion instability is suppressed both for very large and for very small droplets.
Specifically, droplets that are much smaller than the characteristic length of the substrate and product concentration profiles, cannot build up a sufficient difference in the concentrations of substrates and products between the droplet interfaces to sustain self-propulsion.
In the opposing limit where droplets are much larger than the characteristic length of the substrate and product concentration profiles, the diffusion of substrates and products ceases to play a role.
Then, substrates and products reach their local reactive equilibria, and the droplet also cannot build up a sufficient difference in the concentrations of substrates and products between the droplet interfaces to sustain self-propulsion.
In summary, droplet self-propulsion requires the droplet radius to be compatible with the characteristic length of substrate and product concentration profiles, and is optimal for $R \sim l_+$.

\subsection{Quasi-steady state approximation for droplet self-centering}
\label{sec::quasistatic}
\begin{figure}[t]
\centering
\includegraphics{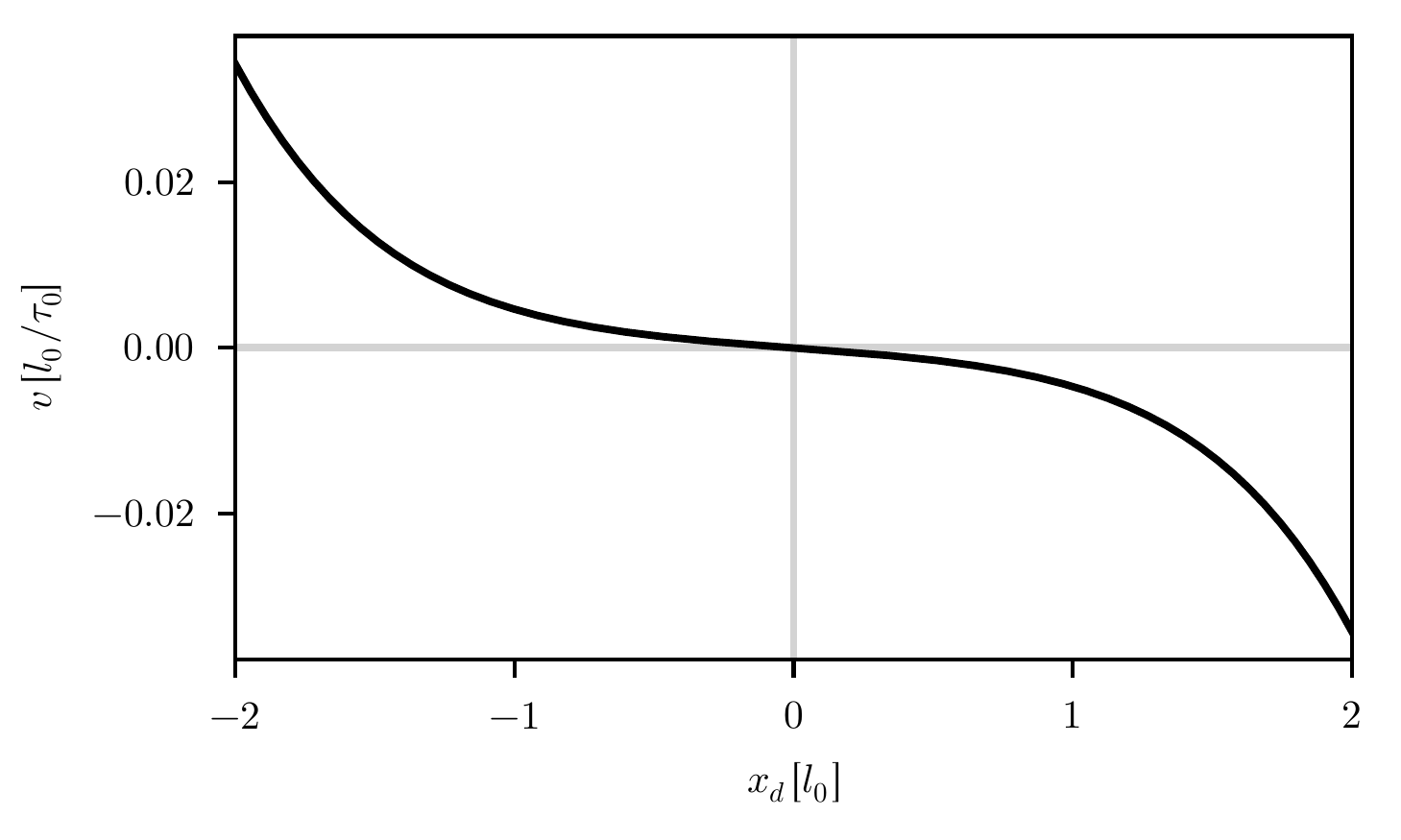}
\caption{%
Droplet velocity as a function of its position in the quasi-steady state approximation, for $M \epsilon_0 / D = 10$. 
The domain has a half-size of ${L = 3 l_0}$, as in Fig.~3b in the main text.
}
\label{fig:quasistatic_velocity}
\end{figure}

From the inequality Eq.~\eqref{eq:motility_condition_full}, we infer that a droplet can self-propel if certain conditions are met, e.g., in the case of large enzyme mobility $M$.
If the droplet exhibits self-propulsion, then it will perform oscillatory motion in a closed domain or adhere to one of the domain boundaries, as discussed in the main text and in section~\fullref{sec:self_centering_simulations}.
However, even if the conditions for self-propulsion are not met, that is when inequality Eq.~\eqref{eq:motility_condition_full} is not fulfilled, Eq.~\eqref{eq:self_consistency} indicates that gradients in the densities of substrates and products will still drive droplet motion.
Such concentration differences between the droplet interfaces can result, for example, from an off-centered position of the droplet in its enclosing domain.
Then, the droplet will gradually position itself towards the domain center.
In the following, we study this scenario theoretically, in the overdamped limit where the droplet does not overshoot past the domain center.
To that end, we consider a quasi-steady state approximation where there is a separation of time scales between the slow motion of the droplet and the fast relaxation of the substrate and product concentration profiles to their pseudo-steady state.
We determine the concentration profiles in this pseudo-steady state using the Python library Sympy~\cite{Meurer2017} for symbolic calculations and confirmed with the computer algebra system Mathematica~\cite{Mathematica}, and provide the full expressions in~\cite{CodeGithub}.
We then calculate the substrate and product concentration values at the droplet interfaces, and insert them into the self-consistency relation~\eqref{eq:self_consistency} to determine the resulting droplet velocity.
This procedure yields the droplet velocity $v(x_d)$ as a function of the position of the droplet center $x_d$, as plotted in Fig.~\ref{fig:quasistatic_velocity}.
Figure~\ref{fig:quasistatic_velocity} indicates that the droplet position has a single stable fixed point at the center of the domain.
Near the domain center, we again use the Python library Sympy~\cite{Meurer2017} and the computer algebra system Mathematica~\cite{Mathematica} to linearize the dynamics and find exponential relaxation with the following timescale:
\begin{equation}
\label{eq:quasistatic_decay_rate}
    \lambda^{-1} = 
    \frac{l_- R \Delta c}{M c_+ \Delta \chi \Delta s^\star} \left[ \cosh\left(\frac{L_\text{free}}{l_-}\right) + \frac{l_+ \sinh\bigl(\frac{L_\text{free}}{l_-}\bigr)}{l_-\tanh\bigl(\frac{R}{l_+}\bigr)} \right] \left[\sinh\left(\frac{L_\text{free}}{l_-}\right) + \frac{l_- \cosh\bigl(\frac{L_\text{free}}{l_-}\bigr)}{l_+ \tanh\bigl(\frac{R}{l_+}\bigr)}\right] \, ,
\end{equation}
where we have defined $L_\text{free} \coloneqq L-R$.
We find a very good agreement between these analytic results and our finite element simulations [Fig.~3c in the main text].

To analyze Eq.~\eqref{eq:quasistatic_decay_rate}, we first discuss a scenario where we keep the value of $\Delta s^\star$ fixed.
Then, Eq.~\eqref{eq:quasistatic_decay_rate} shows that the characteristic time of self-centering diverges for $l_- \ll L_\text{free}$ and for $l_- \gg L_\text{free}$.
Therefore, the typical timescale of finding the domain center is minimal if the characteristic length of the concentration profiles outside of the droplet is comparable to the typical distance towards the domain boundary, $l_- \sim L_\text{free}$.
Furthermore, Eq.~\eqref{eq:quasistatic_decay_rate} also shows that the characteristic time of self-centering diverges for very large droplets $R \gg l_+$ and for very small droplets $R \ll l_+$.
Therefore, droplet self-centering is also optimized if $R \sim l_+$, which allows building up concentration gradients across the droplet.
Consistent with these arguments for the droplet size, for large domain sizes the relaxation rate Eq.~\eqref{eq:quasistatic_decay_rate} scales as
\begin{equation}
    \lambda^{-1} = 
    \frac{\Delta c}{M c_+ \Delta \chi \Delta s^\star} 
    \exp\left(\frac{2L_\text{free}}{l_-}\right) \, \frac{R}{4l_+} \,
    \left[ l_- + \frac{l_+}{\tanh\bigl(\frac{R}{l_+}\bigr)} \right] \left[l_+  + \frac{l_-}{\tanh\bigl(\frac{R}{l_+}\bigr)} \right] \, ,
\end{equation}
which has a minimum as a function of $R/l_+$.

We next discuss how the features of \eqref{eq:quasistatic_decay_rate} relate to Fig.~3c in the main text.
Now, the value of the concentration difference between the reactive equilibria, $\Delta s^\star$, is not fixed because it depends on the catalysis rate $k_1$; see Eq.~\eqref{eq:reactive_equilibria_difference}.
When increasing the catalysis rate $k_1$ to very high values, the characteristic length $l_\pm$ of the concentration profiles becomes very small and one finds that the droplet takes a longer time to find the center of the domain.
In the opposite scenario, where the catalysis rate $k_1$ is very small, one finds that the droplet also takes a longer time to find the domain center because it cannot create significant substrate and product concentration gradients (small value of $\Delta s^\star$).
Therefore, there is an optimal value of the catalysis rate $k_1$ where the droplet finds the center of the domain in the shortest amount of time [Fig.~3c in the main text].

\section{Enzymatic activity of droplets leads to coexistence and arrests the coarsening process}
\label{chapter:coexistence}
So far we have analyzed the dynamics of a single droplet, and have shown how its enzymatic activity can lead to self-propulsion or self-positioning in a finite domain.
Now, we extend our analysis to the dynamics of multiple droplets, in the parameter regime where none of the droplets self-propel.
To that end, we consider small values of the enzyme mobility $M$ where inequality Eq.~\eqref{eq:motility_condition_full} is not fulfilled.
We then analyze under which conditions multiple droplets will show arrested coarsening and therefore coexist.
Before we proceed with this theoretical analysis, we study simulations in the parameter regime where multiple droplets coexist without showing self-propulsion, and analyze how the droplets position themselves in a finite domain.

\subsection{Localization of multiple coexisting droplets}
\label{sec::positioning_multiple}
\begin{figure}[t]
\centering
\includegraphics{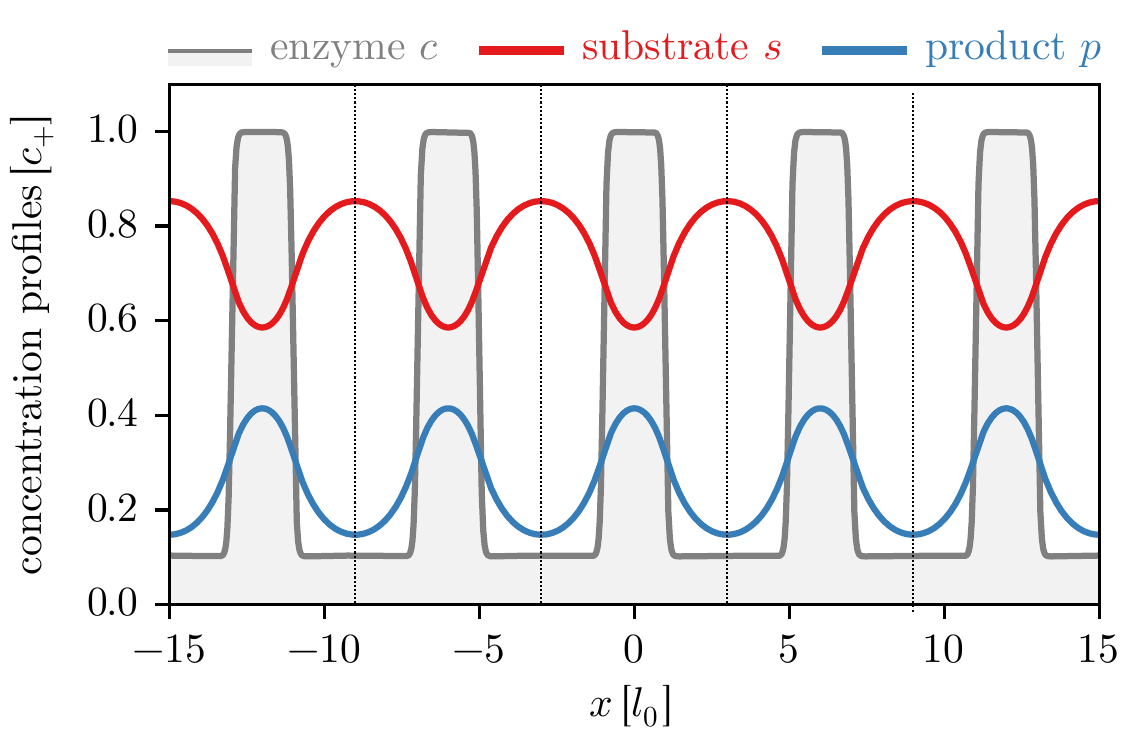}
\caption{%
Stationary concentration profiles corresponding to 5 droplets coexisting on an interval, obtained after simulating the system for a total duration of $1000 \, \tau_0$.
Boundaries of replicate subdomains are indicated by dotted lines. 
As initial condition for the enzyme concentration, we considered 5 distinct droplets of equal radii $R = l_0$.
}
\label{fig:coexistence_1D}
\end{figure}

We have discussed in section~\fullref{sec:self_centering_simulations} that enzymatically active droplets exhibit repulsive interactions with domain boundaries, where enzymatic substrate depletion is enhanced.
Analogously, two neighboring droplets will strongly deplete the substrate in the region between them.
The resulting substrate concentration gradients will, through attractive enzyme--substrate interactions, drive enzyme fluxes within each droplet that lead away from the neighboring droplet.
This leads to effective repulsive interactions among droplets, which suggest that droplets will position themselves equidistantly in a finite domain.
For a steady state in which $N$ droplets coexist in the same domain without showing self-propulsion [Fig.~\ref{fig:coexistence_1D}], equidistant positioning indicates a replica symmetry in which the domain can be divided into a chain of $N$ identical subdomains, each containing only one droplet [Fig.~\ref{fig:coexistence_1D}].
In such a steady state, there is no net particle exchange across the interfaces between adjacent subdomains, which is equivalent to no-flux boundary conditions at each subdomain boundary. 
Hence, in each individual subdomain, the droplet will localize to the center according to sections~\fullref{sec:self_centering_simulations}~and~\fullref{sec::quasistatic}.
Because of this symmetry of replicate subdomains, the distance between one of the domain boundaries and the nearest droplet, $L/N$ for a domain of size $2L$, will be exactly half of the distance between two adjacent droplets, $2L/N$.
In the next section, we illustrate why enzymatic droplets can coexist in the first place.

\subsection{Coexistence of droplets in a one-dimensional domain}
\label{sec::coexistence_1D}
\begin{figure}[t]
\centering
\includegraphics{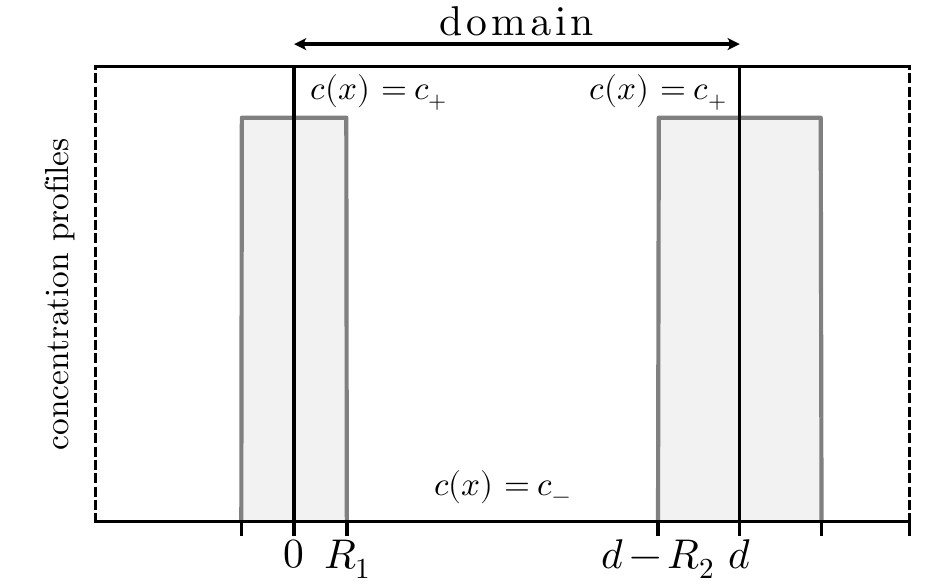}
\caption{%
Illustration of the sub-domains in the sharp-interface approximation, for two droplets that gradually equilibrate their sizes.
Gray areas illustrate subdomains, as discussed in Fig.~\ref{fig:illustration_helmholtz}.
In each subdomain, we only need to solve a Helmholtz equation. 
To simplify this mathematical problem, we construct a scenario where the substrate and product concentration profiles are approximately symmetric with respect to the droplet centers. 
Specifically, for our analytical calculations we consider a domain that ends at the droplet centers (black vertical lines) and has no-flux boundary conditions there.
This construction is exact if, for an alternating chain of large and small droplets, the substrate and product concentration profiles are symmetric with respect to each droplet center (replica symmetry).
In our simulations, this would correspond to two droplets that are positioned equidistantly in a simulation domain with periodic boundary conditions at the position indicated by the dashed lines.
For simulation domains with no-flux boundary conditions at the position indicated by the dashed lines, our simplification is nevertheless a good approximation.
}
\label{fig:helmholtz_coexistence}
\end{figure}

To illustrate the mechanism underlying the coexistence of multiple droplets, we study a scenario with only two droplets of differing sizes $R_1$ and $R_2\geq R_1$ [Fig.~\ref{fig:helmholtz_coexistence}].
The larger droplet depletes more substrate and accumulates more product than the smaller droplet, which can be illustrated with the following limiting cases:
For droplet sizes much larger than the characteristic length of the substrate and product concentration profiles inside of the droplet, $R_2 \gg l_+$, the substrate and product concentrations at the droplet boundaries are given by their reactive equilibria%
\footnote{The reactive equilibria correspond to the solution of the reaction-diffusion equations for substrates and products, Eq.~\eqref{eq:substrates}~and~\eqref{eq:products}, in the limit $D\rightarrow 0$ and $\Lambda \rightarrow 0$.}.
For droplet sizes much smaller than the characteristic length of the concentration profiles inside of the droplet, $R_1 \ll l_+$, diffusion will homogenize the substrate and product concentration profiles. 
Then, the concentrations at the droplet boundaries remain at their ambient concentration values in the far field. 
The resulting concentration gradients will, because of strongly attractive enzyme--substrate interactions and weaker enzyme--product interactions, drive a net flux of enzymes from the larger to the smaller droplet, see Eq.~\eqref{eq:enzymes}.
Therefore, the size difference between the two droplets, $\Delta R = R_2 - R_1$ will shrink over time, indicating coexistence [Supplemental Video~\ref{vid:coexistence}].

In the following, we derive the characteristic timescale of droplet size equilibration.
For droplets that do not exhibit self-propulsion, the dynamics of enzymes is very slow compared to the dynamics of substrates and products.
Therefore, in full analogy to section~\fullref{sec::quasistatic}, we make a quasi-steady state approximation for the fast relaxation of the substrate and product concentration profiles in response to the quasi-static enzyme concentration profile:
\begin{equation}
c(x)=
\begin{cases} 
c_+ & \text{for } \| x \| \leq R_1 \quad \text{and} \quad \| x - d \| \leq R_2\, , \\
c_- & \text{otherwise},
\end{cases}
\end{equation}
where the droplet centers are located at $x=0$ and $x=d$.
Given this concentration profile of enzymes, in the scenario with nonreciprocal interactions ($\Lambda = 0$) one can then analytically solve Eq.~\eqref{eq:stationarysubstrate} for the concentration profiles of substrates $s(x)$ and products $p(x) = n - s(x)$.
This mathematical problem is practically identical to sections~\fullref{sec::sharp_interface}~and~\fullref{sec::quasistatic}, as illustrated in Fig.~\ref{fig:helmholtz_coexistence}, and the full expressions are provided in~\cite{CodeGithub}.
Here, we outline the main assumptions and simplifications of the derivation.

In principle, for droplets that are close to each other, one observes a net substrate and product concentration gradient across each individual droplet.
This net concentration gradient will lead to an effective repulsion between the two droplets, analogous to sections~\fullref{sec:self_centering_simulations}~and~\fullref{sec::positioning_multiple}, and prevent coalescence.
Here, however, we are interested in the particle exchange currents between the two droplets. 
Therefore, we focus on the substrate and product concentration differences between the nearest interfaces of the two \emph{different} droplets, $\Delta s \coloneqq s(d-R_2) - s(R_1)$ and $\Delta p \coloneqq p(d-R_2) - p(R_1)$.
To simplify our analysis, we construct a scenario where the substrate and product concentration profiles are approximately symmetric with respect to each individual droplet, so that each droplet will remain immobile as discussed in section~\fullref{sec::positioning_multiple}.
Specifically, we consider no-flux (reflective) boundary conditions at the droplet centers [Fig.~\ref{fig:helmholtz_coexistence}], which would be exact for a periodic chain of big and small droplets.
Then, we solve for the concentration profiles of substrates and products in the domain $[0, d]$, and find the substrate concentration difference between the closest interfaces of the two droplets~\cite{CodeGithub},
\begin{equation}
\label{eq:coexistance_imbalance}
    \Delta s = - \Delta s^\star \frac{\Delta R}{l_+} 
    \left[ \sinh\left(\frac{R}{l_+}\right) + \frac{l_+ \, \cosh\bigl(\frac{R}{l_+}\bigr)}{l_- \, \tanh\bigl(\frac{L_\text{free}}{l_-}\bigr)} \right]^{-1}
    \left[ \cosh\left(\frac{R}{l_+}\right) + \frac{l_- \, \sinh\bigl(\frac{R}{l_+}\bigr)}{l_+ \, \tanh\bigl(\frac{L_\text{free}}{l_-}\bigr)} \right]^{-1} \, ,
\end{equation}
for small differences in the droplet radii $\Delta R \coloneqq R_2 - R_1$.
Here, we have defined the mean droplet radius, $R\coloneqq (R_1+R_2)/2$, and the free half-distance between droplets $L_\text{free} \coloneqq d/2 - R$.

These concentration differences drive an exchange current of enzymes between the two droplets, which we in analogy to \fullref{sec::induced_flux} determine by integrating the continuity equation Eq.~\eqref{eq:enzymes} over the inter-droplet domain $[R_1, d - R_2]$,
\begin{equation}
\label{eq:exchange_flux_1d}
    j_\times \, 2 L_\text{free} = - M c_- \, \bigl[ \chi_s \, \Delta s + \chi_p \, \Delta p \bigr] \, ,
\end{equation}
where we have omitted the contribution of the Cahn-Hilliard chemical potential $\mu_0(z)$ as explained next.

\begin{figure}[t]
\centering
\includegraphics{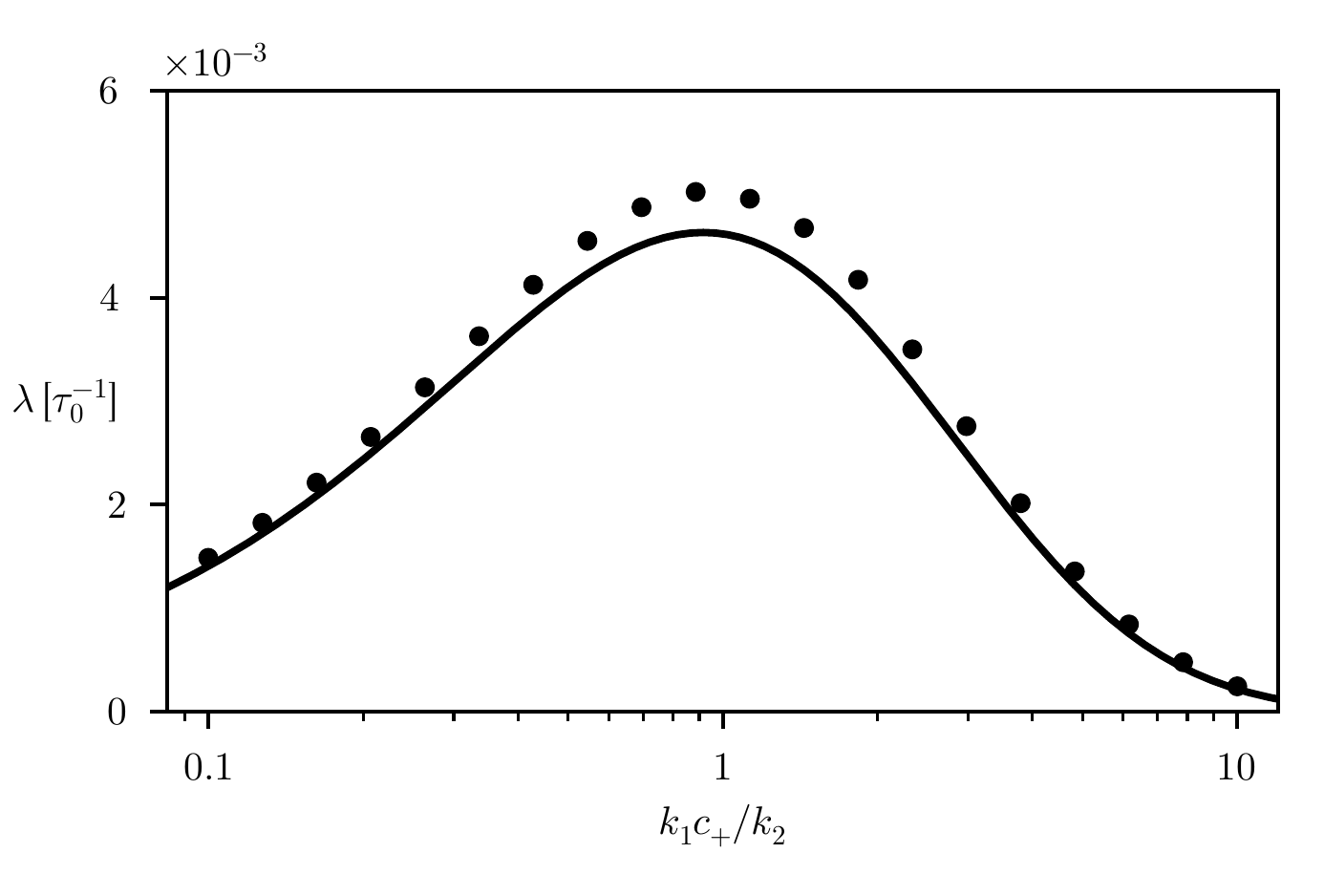}
\caption{%
Droplet size equilibration rate as a function of the reaction rate $k_1$. 
The solid line indicates the expression obtained in the quasistatic limit~\eqref{eq:decay_rate_radius}, and agrees well with the results of numerical simulations (dots). 
The droplets are initially positioned at $x_1(0) = -2.5 l_0$ and $x_2(0) = 2.5 l_0$ in the domain $[-5l_0, 5l_0]$, have radii $R_1(0) = 0.9 l_0$ and $R_2(0) = 1.1 l_0$, and have an interface width of ${w = 0.05 l_0}$.
}
\label{fig:radius_difference_decay_rate}
\end{figure}

In general, the interaction-driven exchange flux~\eqref{eq:exchange_flux_1d} will be superimposed by a coarsening current that stems from the Cahn-Hilliard chemical potential $\mu_0(z)$, which drives slow coarsening in one dimension.
In the sharp-interface limit and in 1d, however, coarsening becomes infinitely slow.
Therefore, in 1d the flux given by Eq.~\eqref{eq:exchange_flux_1d} can easily drive a dynamics that is opposite to coarsening, by transporting material from the larger droplet to the smaller droplet.
The size difference between the two droplets then gradually changes with time, and is in a finite 1d domain with no-flux boundary conditions determined by%
\footnote{In this case, the flux is given by $\Delta c \, \partial_t (2R_2) = - \Delta c \, \partial_t ( 2R_1) = j_\times$. For periodic domains, an additional factor of 2 would enter because one has particle exchange with twice as many droplets.}
$j_\times = \Delta c \, \bigl[ \partial_t (R_2 - R_1) \bigr] =  \Delta c \, \partial_t \Delta R$.
Taken together, the size difference between the two droplets will evolve as follows:
\begin{equation}
\label{eq:coexistence_dynamics}
    \partial_t \Delta R = \frac{M \, c_- \, \Delta\chi}{2 \, L_\text{free} \, \Delta c} \, \Delta s \, ,
\end{equation}
where we have used ${\Delta p = - \Delta s}$ and defined ${\Delta\chi \coloneqq \chi_p - \chi_s}$.
After inserting Eq.~\eqref{eq:coexistance_imbalance} into Eq.~\eqref{eq:coexistence_dynamics}, one finds that the size difference decays exponentially with a characteristic timescale
\begin{equation}
\label{eq:decay_rate_radius}
    \lambda^{-1} = \frac{2 \, L_\text{free} \, l_+ \, \Delta c}{M \, c_- \, \Delta\chi \, \Delta s^\star} \,
    \left[ \sinh\left(\frac{R}{l_+}\right) + \frac{l_+ \, \cosh\bigl(\frac{R}{l_+}\bigr)}{l_- \, \tanh\bigl(\frac{L_\text{free}}{l_-}\bigr)} \right]
    \left[ \cosh\left(\frac{R}{l_+}\right) + \frac{l_- \, \sinh\bigl(\frac{R}{l_+}\bigr)}{l_+ \, \tanh\bigl(\frac{L_\text{free}}{l_-}\bigr)} \right] \, .
\end{equation}
This result is equivalent to Eq.~\eqref{eq:quasistatic_decay_rate} under the replacements $c_+ \leftrightarrow c_-$, $l_+ \leftrightarrow l_-$, and $L_\text{free} \leftrightarrow R$, because it corresponds to the self-centering of an inverted droplet\footnote{In comparison to Eq.~\eqref{eq:quasistatic_decay_rate}, there is an additional factor of $2$. This factor would drop out in a periodic domain, where droplet size equilibration is twice as fast because each droplet can exchange particles with two neighbors. 
}
Thus, our discussion for the self-centering of droplets, see section~\fullref{sec::quasistatic}, also applies here.
In particular, droplet size equilibration is fastest when the length scales of the droplets, concentration profiles of substrates and products, and the distance between droplets are compatible.
Comparing our analytical results to simulations, we find good agreement [Fig.~\ref{fig:radius_difference_decay_rate}].

\subsection{Coexistence of droplets in a three-dimensional domain}
\label{sec::coexistence_3D}
In the previous section, we have seen that enzyme--substrate and enzyme--product interactions can easily oppose and reverse the coarsening process in 1d, which becomes infinitely slow in the sharp-interface limit in 1d.
For 2d and 3d droplets, an additional effect comes into play that arises from the \emph{curvature} of the droplet interfaces: the Laplace pressure leads to an evaporation of small droplets and stabilization of large droplets, thus greatly accelerating the coarsening process~\cite{LifshitzSlyozov1961, Wagner1961, Bray1994}.
This means that in 2d and 3d the interaction-driven exchange currents between different droplets are superimposed by much stronger coarsening currents than in 1d, which narrows down the parameter regime in which one can observe arrested phase separation.
In the following, we derive a criterion to still observe arrested coarsening in 2d and 3d.

We consider a droplet with radius $R$ and small interface width ${w=\sqrt{2\kappa/r} \ll R}$, which has the homogeneous free energy density $f(c) = \frac{u}{4}(c-\tilde c)^4 - \frac{r}{2}(c-\tilde c)^2$; see Eq.~\eqref{eq:freeenergy_cahnhilliard}.
The surface tension of this droplet is given by $\gamma = 2 r^2 w / (3 u)$ and induces a Laplace pressure, which leads to an increase in the enzyme concentration just outside of the droplet as described by the Gibbs-Thomson relation~\cite{Bray1994, Weber2019}:
\begin{equation} 
\label{eq:gibbs_thomson}
\delta c(R) = \frac{2\gamma}{R} \frac{1}{\Delta c \, f''(c_-)} = \frac{1}{6} \frac{w}{R} \, \Delta c \, ,
\end{equation}
where $f''(c_-) = 2 r$.
The increase in the enzyme concentration affects the Cahn-Hilliard chemical potential outside of the droplet, $\mu_0 \simeq 2 r \, \delta c$, and is more pronounced for small droplets, see Eq.~\eqref{eq:gibbs_thomson}.
This leads to a difference in the Cahn-Hilliard chemical potential between the two droplets, $\Delta\mu_0$, which drives diffusive transport of enzymes from small towards large droplets:
\begin{equation}
\label{eq:exchange_flux_3d_coarsening}
    j_0 \approx -M c_- \, \frac{\Delta\mu_0}{d} \approx -2 \, r \, M c_- \, \frac{\delta c(R_2) - \delta c(R_1)}{d} = \frac{r \, M c_- w \, \Delta c}{3 \, R_1\, R_2 \, d} \, \Delta R \, ,
\end{equation}
where we have assumed that the droplets are far apart, $d \gg R_{1,2}$, and defined $\Delta R \coloneqq R_2 - R_1$.

\begin{figure}[t]
\centering
\includegraphics{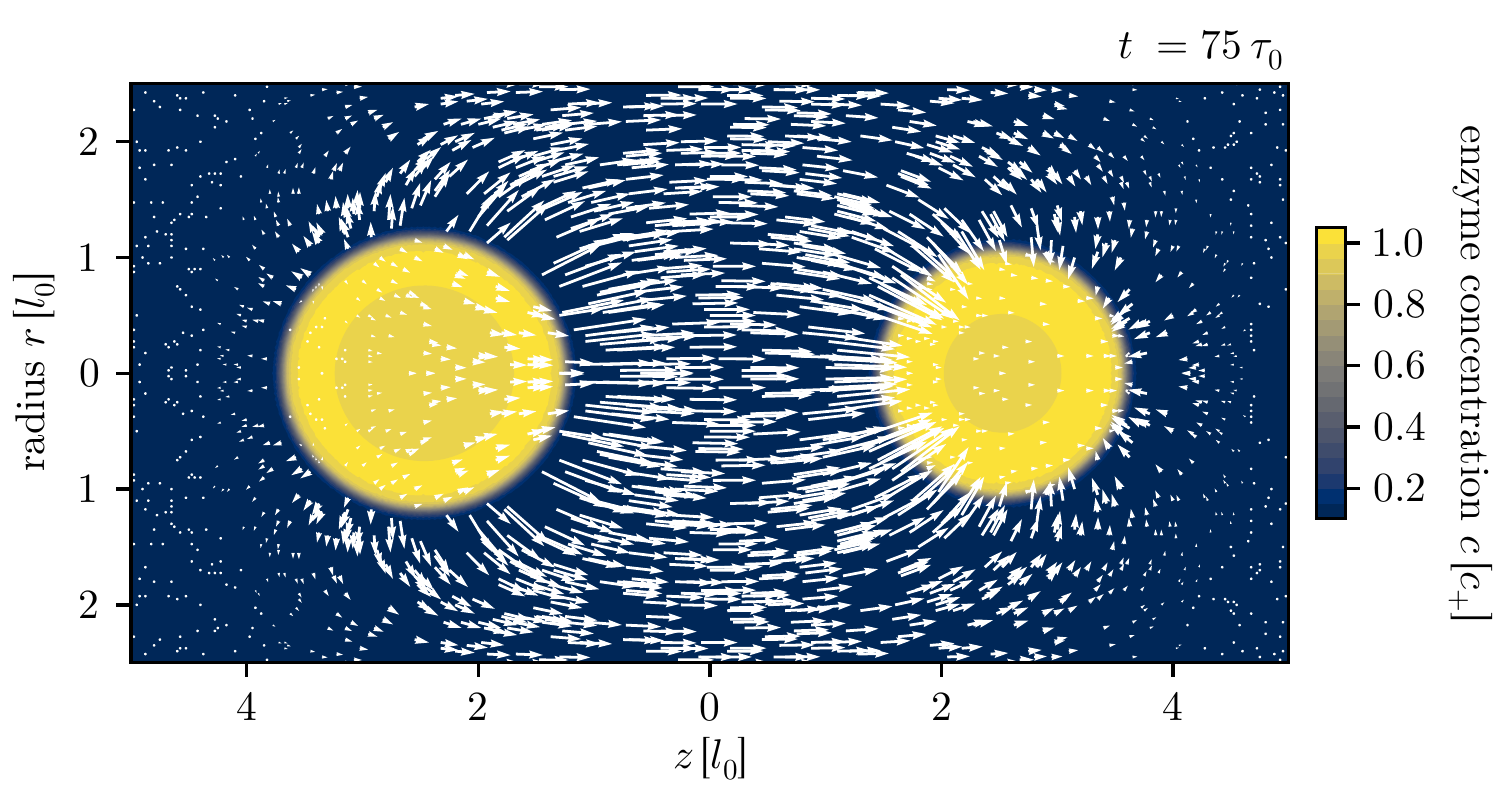}
\caption{%
Snapshot of a simulation of two coexisting droplets in 3d. 
The color map indicates the local enzyme concentration, and the white arrows are proportional to the local mean velocity of the enzymes $\boldsymbol{j}_c(\boldsymbol{x})/c(\boldsymbol{x})$.
The simulation was performed in a rotationally invariant cylindrical domain of radius ${L_r = 2.5 l_0}$ and half-height ${L_z = 5 l_0}$.
Parameters: ${M \epsilon_0 / D = 10}$, ${w = 0.05 l_0}$ as in Fig.~4a in the main text, with $k_1 c_+/k_2 = 1$ and $\chi_s / r = -0.21$.
}
\label{fig:local_velocities_coexistence}
\end{figure}

As discussed in section~\fullref{sec::coexistence_1D}, the enzymatic activity of the droplets locally depletes substrate and accumulates product, which results in substrate and product concentration gradients that can contribute to the exchange flux between different droplets, see Eq.~\eqref{eq:exchange_flux_1d}.
For droplets that are far apart, $d \gg R_{1,2}$, one then has
\begin{equation}
\label{eq:exchange_flux_3d_interactions}
    j_\times \approx \frac{M \, c_- \, \Delta \chi}{d} \, \Delta s \, ,
\end{equation}
where we have again used $\Delta p = - \Delta s$, and defined $\Delta\chi \coloneqq \chi_p - \chi_s$.
If the droplets are far apart, then we can determine the local concentration profiles of substrates and products independently for each droplet, by solving a Helmholtz equation in spherical coordinates\footnote{A general analytic solution for droplets that are close to each other would require solving the Helmholtz equation in bispherical coordinates. However, the Helmholtz equation is not separable in bispherical coordinates.}:
\begin{equation}
\label{eq:stationarysubstrate_3D_radial}
    \frac{D}{r^2} \, \partial_r \bigl[ r^2 \partial_r s(r) \bigr] - \bigl[k_1 c(r) + k_2 \bigr] \, s(r) + k_2 n = 0 \, ,
\end{equation}
where we make a sharp-interface approximation for the concentration of enzymes, $c(r) = c_\text{sharp}(r)$, see Eq.~\eqref{eq:enzyme_concentration_sharp}.
We determine the substrate and product concentration profiles as described in section~\fullref{sec::sharp_interface}; the full expressions are provided in~\cite{CodeGithub}.
The difference between the substrate concentration values at the interfaces of two droplets with a small difference in size, $\Delta R \ll l_+$, are then given by:
\begin{equation}
\label{eq:substratedifference_3D}
    \Delta s \approx - \Delta s^\star \, \frac{ l_+^{-1} [ \sinh(2  R/l_+) - 2 R/l_+] + 2 l_-^{-1} [\sinh^2(R/l_+) - (R/l_+)^2]}{2 R^2 \left[ l_+^{-1} \cosh(R/l_+) + l_-^{-1} \sinh(R/l_+) \right]^2} \Delta R \, .
\end{equation}
As discussed in section~\fullref{sec::coexistence_1D}, the larger droplet typically depletes more substrate, leading to a substrate concentration difference proportional to the size difference: $\Delta s \propto - \Delta R$.
This concentration difference drives a flux of enzymes from the larger to the smaller droplet [Fig.~\ref{fig:local_velocities_coexistence}].
We now substitute Eq.~\eqref{eq:substratedifference_3D} into the expression for the interaction-driven exchange fluxes that oppose coarsening, Eq.~\eqref{eq:exchange_flux_3d_interactions}, and ask under which conditions these fluxes may dominate ($j_\times + j_0 < 0$) over the coarsening currents given by Eq.~\eqref{eq:exchange_flux_3d_coarsening}.
This comparison yields a criterion to observe arrested coarsening, which is fulfilled if the interactions are sufficiently strong:
\begin{equation}
\label{eq:critical_interaction_parameter}
\Delta \chi \gtrsim
    \frac{2 r w \Delta c \left[ l_+^{-1} \cosh(\bar R/l_+) + l_-^{-1}  \sinh(\bar R/l_+) \right]^2}{3 \Delta s^\star \left[l_+^{-1}  ( \sinh(2 \bar R/l_+) - 2 \bar R/l_+) + 2 l_-^{-1}  (\sinh^2(\bar R/l_+) - (\bar R/l_+)^2) \right]} \,.
\end{equation}
Comparing this analytical criterion with numerical simulations, we find good agreement [Fig.~4a in the main text, dynamics shown in Supplemental Video~\ref{vid:coexistence}].

\subsection{Droplet division}
\label{sec::division}

\begin{figure}[hbtp]
\centering
\includegraphics{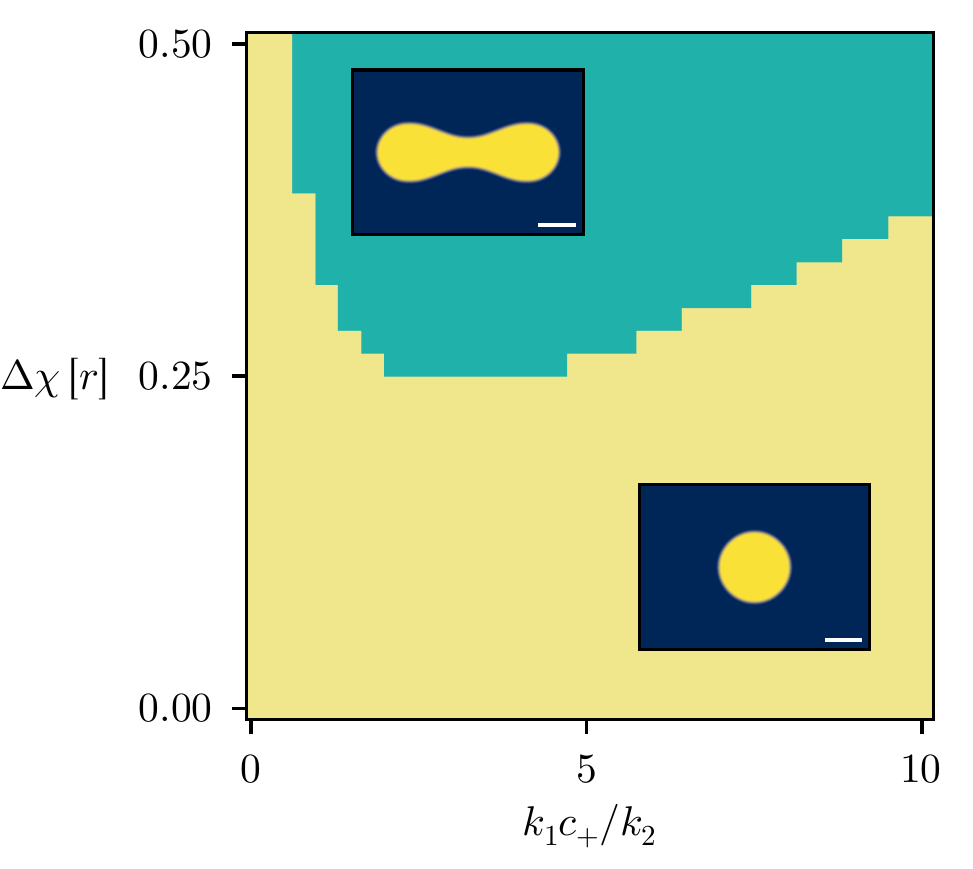}
\caption{%
Results of numerical simulations starting with a single droplet.  The droplet can either remain spherical (yellow regime) or start elongating (cyan regime) depending on the values of $k_1$ and $\Delta \chi$. 
The simulation was performed in a rotationally invariant cylindrical domain of radius ${L_r = 2.5 l_0}$ and half-height ${L_z = 3.5 l_0}$.
Parameters: ${M \epsilon_0 / D = 10}$, and ${w = 0.05 l_0}$ as in Fig.~4b in the main text.}
\label{fig:elongation}
\end{figure}

We have seen that the enzymatic activity of the droplet, coupled with enzyme--substrate and enzyme--product interactions, can oppose and stop coarsening.
Then, could this mechanism also lead to a shape instability and to divisions of droplets? 
In fact, in 2d and 3d, substrate depletion is enhanced for smaller curvatures of the droplet interface and reduced for larger curvatures of the droplet interface, by having a smaller interface over which substrate can be resupplied from the environment relative to the enclosed volume where substrate is depleted by enzymes.
This effect is analogous to the stronger substrate depletion by larger spherical droplets when compared to smaller droplets, and can drive a net flux of enzymes from regions with a small curvature of the droplet interface towards regions with a large curvature of the droplet interface.
Regions with a large curvature will then move outwards while regions with a small curvature will move inwards, elongating the droplet and further increasing the differences in curvature (positive feedback mechanism), while the droplet volume remains conserved.
We observed such droplet elongation in simulations for sufficiently large values of the interaction parameter $\Delta \chi$ [Fig.~\ref{fig:elongation}].

Supplemental Video~\ref{vid:division} shows a simulation of a controlled division process in 3d.
We take the catalysis rate $k_1$ of converting products into substrates as a control parameter. 
We start with a large value of $k_1$, for which a single droplet is stable at the center of the domain.
Then, upon lowering $k_1$, we observe the following dynamics.
The droplet first elongates (initial shape instability) and then forms a dumbbell shape with a neck.
The tube connecting the two dumbbells gradually becomes thinner until it pinches off, leaving two separated droplets (division).
These two droplets assume a spherical shape and position themselves equidistantly in the domain as expected from the discussion of section \ref{sec::positioning_multiple}.
If we increase $k_1$ to its initial value, then the two droplets remain stable, showing that the process is irreversible.
Figure~4b in the main text shows some snapshots of the enzyme concentration, with $t=0$ being the time when $k_1$ is switched to a smaller value.

Note that we have only observed droplet divisions in 3d, but never in 2d or 1d.
This suggests an effect that destabilizes an elongated cylindrical shape in favor of spherical shapes.
Such an effect is characteristic for classical pearling instabilities driven by surface tension~\cite{Rayleigh1878}, where spherical shapes have smaller surface area than cylindrical shapes with the same volume.
Therefore, cell division should be controlled by the interface width $w$, which not only sets the value of the surface tension, but also defines a length scale over which the two sides of a thin cylinder that connects the two dumbbells can interact.

\section{Multi-component condensates}
So far, we have studied a scenario where the enzymes undergo spontaneous phase separation on their own.
However, one can also envision a much more general scenario where the enzymes do not phase separate spontaneously, but are only enriched in a droplet that consists of a scaffold protein with concentration $q(\mathbf{x})$.
In this scenario, the scaffold proteins spontaneously phase separate driven by the chemical potential $\mu_0 (q)$ of the Cahn-Hilliard model, and only interact with the enzymes through a Flory-Huggins coupling $\chi_q$:
\begin{equation}
    \partial_t q = \boldsymbol\nabla \cdot \Bigl\{ M_q q \boldsymbol\nabla \bigl[\mu_0(q) + \chi_q \, c \bigr]\Bigr\} \,. 
\label{eq:multi_scaffold_proteins}
\end{equation}
We assume that the enzymes are present at relatively small concentrations so that they show currents driven by diffusion, and effective Flory-Huggins couplings to the scaffold proteins, substrates, and products:
\begin{equation}
    \partial_t c = \boldsymbol\nabla \cdot \Bigl\{ D_c \boldsymbol\nabla c + M_c c \boldsymbol\nabla \bigl[ \chi_q q + \chi_s s + \chi_p p \bigr] \Bigr\} \, .
\label{eq:multi_enzymes}
\end{equation}
Attractive effective interactions between the scaffold proteins and the enzymes, $\chi_q < 0$, lead to an enrichment of enzymes in the droplet.
In the sharp-interface limit, this enrichment is quantified by
\begin{equation}
    \frac{c_+}{c_-} = \exp\left[-\frac{M_c}{D_c} \chi_q \left(q_+ - q_-\right)\right]\, ,
\end{equation}
where the indices $\pm$ indicate the concentrations on the inner and the outer side of each droplet interface, respectively.
Analogous to droplets that consist mainly of enzymes, which we have discussed so far, our simulations show self-propulsion if the mobilities of the scaffold proteins ($M_q$) and enzymes ($M_c$) are sufficiently large, and self-centering otherwise (Supplemental Video~\ref{vid:multicomponent}).
The enzymes then act as a link, by mediating the spatial organization of substrate and product through the droplet, as well as transmitting forces to the droplet that arise due to these inhomogeneous substrate and product concentration profiles.

\clearpage\newpage

\section{Supplemental Videos}

In the following, we describe the videos available as supplemental material.

\begin{enumerate}

\item \path{video_1_self_propulsion_instability.mp4}. 
\label{vid:self_propulsion}
For all simulations, we set $M = 1000 D/\epsilon_0$. 
\emph{1d droplet:} Evolution of the concentration profiles of enzymes, substrates and products resulting from a numerical simulation of Eqs.~\eqref{eq:system} in a one-dimensional interval with no-flux boundary conditions and half-length $L = 10 l_0$.  As initial condition we consider a single droplet of enzymes at the center of the interval, the starting concentrations of substrates and products are the equilibrium values of the reaction terms plus small random perturbations in the droplet region.  We observe the self-propulsion instability, the droplet starts moving in a random direction determined by the initial conditions.
\emph{2d droplet:} Evolution of the concentration profile of enzymes resulting from a numerical simulation performed in a two-dimensional circular domain of radius $L_r = 7 l_0$.  
\emph{3d droplet:} Evolution of the concentration profile of enzymes resulting from a numerical simulation performed in a three-dimensional cylindrical domain of radius $L_r = 4 l_0$ and half-height $L_z = 7 l_0$.

\item \path{video_2_positioning_1d.mp4}. 
\label{vid:positioning_1d}
Evolution of the concentration profiles of enzymes, substrates and products resulting from numerical simulations analogous to the one for the 1d droplet in Supplemental Video~\ref{vid:self_propulsion}.  As initial condition we consider a single droplet of enzymes positioned at $x_d(0) = -l_0$, the starting concentrations of substrates and products are the stationary profiles that they would reach in the absence of interactions.  The parameter values are the same as for Fig.~3a in the main text.

\item \path{video_3_positioning_2d_and_3d.mp4}.
\label{vid:positioning_2d_3d}
\emph{2d droplet:} Evolution of the concentration profiles of enzymes resulting from a numerical simulation performed in a two-dimensional square domain of half-side length $L = 3 l_0$.  As initial condition we consider a single droplet of enzymes positioned at $(x,y)=(-l_0,0)$, the starting concentrations of substrates and products are the equilibrium values of the reaction terms.  The droplet moves to the center of the domain and localizes there.  The other parameter values are the same as for Supplemental Video~\ref{vid:positioning_1d}.  
\emph{3d droplet:} Evolution of the concentration profiles of enzymes resulting from a numerical simulation performed in a three-dimensional cylindrical domain of radius $L_r = 3 l_0$ and half-height $L_z = 3 l_0$.  The droplet is initially positioned at $z = -l_0$.

\item \path{video_4_coexistence.mp4}.  
\label{vid:coexistence}
\emph{1d droplets:} Evolution of the concentration profiles of enzymes, substrates and products resulting from a numerical simulation analogous to the one for the 1d droplet in Supplemental Video~\ref{vid:self_propulsion}.  As initial condition we consider two distinct droplets of enzymes of radii $R_1(0) = 0.5 l_0$ and $R_2(0) = 1.5 l_0$ positioned at $x=\mp 2.5 l_0$.  The starting concentrations of substrates and products are the equilibrium values of the reaction terms.  Enzymes are transported from the larger droplet to the smaller one until the radii of the two droplets become equal.  
\emph{3d droplets:} Evolution of the concentration profiles of enzymes resulting from a numerical simulation analogous to the one for the 3d droplet in Supplemental Video~\ref{vid:self_propulsion}.  The initial conditions are analogous to the one for the 1d droplets.  $L_r = 2.5 l_0$, $L_z = 5 l_0$, $\chi_s / r = -0.21$, $w = 0.05 l_0$.

\item \path{video_5_droplet_division.mp4}. 
\label{vid:division}
Evolution of the concentration profiles of enzymes resulting from a numerical simulation performed in a three-dimensional cylindrical domain of radius $L_r = 1.5 l_0$ and half-height $L_z = 4 l_0$.  As initial condition we considered a droplet of enzymes positioned at the center of the domain.  The starting concentrations of substrates and products are the equilibrium values of the reaction terms.  The simulation starts with a catalysis rate $k_1 = 100 \, k_2/c_+$ for which the droplet is stable.  Then the catalysis rate is switched to $k_1 = 1 \, k_2/c_+$ and the droplet divides into two.  Finally, the catalysis rate is switched again to $k_1 = 100 \, k_2/c_+$ and the two droplets maintain their stability. Parameters: $M = 10 D/\epsilon_0$, $\chi_s = -0.5 r$, $w = 0.05 l_0$.

\item \path{video_6_multicomponent_droplets.mp4}.
\label{vid:multicomponent}
Evolution of the concentration profiles of scaffold proteins, enzymes, substrates and products resulting from numerical simulations of Eqs.~(\ref{eq:substrates}, \ref{eq:products}, \ref{eq:multi_scaffold_proteins}, \ref{eq:multi_enzymes}).  
We consider as initial condition a droplet of scaffold proteins and a uniform concentration of enzymes $c(t=0)=0.25 q_+$.
\emph{Self-propulsion:} Simulation analogous to the 1d droplet shown in Supplemental Video~\ref{vid:self_propulsion}. 
Parameters: $\chi_q / r = -0.2$, $M_q = M_c = 1000 D/\epsilon_0$, $D_c/D = 100$.
\emph{Self-centering:} Simulation analogous to Supplemental Video~\ref{vid:positioning_1d}. 
Parameters: $\chi_q / r = -0.2$, $M_q = M_c = 100 D/\epsilon_0$, $D_c/D = 10$.

\end{enumerate}

\clearpage\newpage

\bibliography{droplets}